\documentclass[sigconf,screen]{acmart}

\RequirePackage{OpenProc-Mods}

\AddToHook{cmd/maketitle/before}{%

\def\EDBTISSN{2367-2005}
\def\EDBTISBN{978-3-89318-106-3}
\setcopyright{none}
\copyrightyear{\copyright\ 2026 Copyright held by the owner/author(s).
  Published on \url{OpenProceedings.org} under ISBN \EDBTISBN, series ISSN \EDBTISSN.
  Distribution of this paper is permitted under the terms of the
  Creative Commons license CC-by-nc-nd 4.0}
\acmYear{2026}
\acmDOI{}
\acmISBN{}

\acmConference[EDBT '27]{Extending Database Technology}{6-9 April 2027}{Lille (France)}

\settopmatter{printacmref=false, printccs=false, printfolios=false}

\newsavebox{\ximagebox}
\newlength{\ximageheight}
\newsavebox{\xglyphbox}
\newlength{\xglyphheight}
\newcommand{\xbox}[1]%
  {\savebox{\ximagebox}{#1}%
  \settoheight{\ximageheight}{\usebox{\ximagebox}}%
  \savebox{\xglyphbox}{\color{white}\char32}%
  \settoheight{\xglyphheight}{\usebox{\xglyphbox}}%
  \raisebox{\ximageheight}[0pt][0pt]{\raisebox{-\xglyphheight}[0pt][0pt]{%
    \makebox[0pt][l]{\usebox{\xglyphbox}}}}%
    \usebox{\ximagebox}%
    \raisebox{0pt}[0pt][0pt]{\makebox[0pt][r]{\usebox{\xglyphbox}}}}

\newsavebox{\LogoBox}
\sbox{\LogoBox}{\includegraphics[height=1cm]{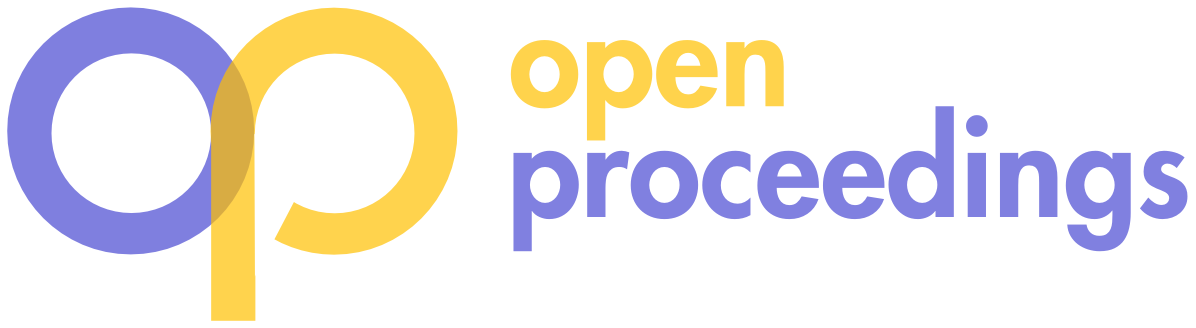}}

\fancypagestyle{firstpagestyle}{%
  \fancyhf{}%
  \fancyfoot{}%
  \fancyhead[L,C]{}
  \fancyhead[R]{\href{https://OpenProceedings.org/}{\raisebox{15pt}[0pt][0pt]{\xbox{\usebox{\LogoBox}}}}}
  }
  
\hypersetup{%
  pdfcopyright={CC-by-nc-nd},
  pdfpublisher={OpenProceedings.org -- University of Konstanz, Germany},
  pdfcreator={LaTeX with acmart, hyperref, and EDBT modifications},
  pdfvolumenum={30},
  pdfissuenum={1},
  pdfisbn={\EDBTISBN},
  pdfissn={\EDBTISSN},
  pdfdoi={10.48786/edbt.2027.xx}
}
}

\def\BibTeX{{\rm B\kern-.05em{\sc i\kern-.025em b}\kern-.08em
    T\kern-.1667em\lower.7ex\hbox{E}\kern-.125emX}}

\usepackage[nameinlink]{cleveref}
\usepackage{listings}
\usepackage{array}
\usepackage{colortbl} 
\usepackage{subcaption} 
\usepackage{tcolorbox} 
\definecolor{darkred}{rgb}{139, 0, 0}

\newcommand*\numcircledmod[1]{\raisebox{.5pt}{\textcircled{\fontfamily{lmss}\selectfont\footnotesize\raisebox{-.4pt} {#1}}}}
\newtcolorbox[auto counter]{guideline}[1][]{
    width=\linewidth,
    top=0pt,bottom=0pt,left=0pt,right=0pt,boxsep=2pt,
    before skip=1ex,after skip=1ex,left skip=0pt,
    boxrule=1pt, #1
}

\begin{document}


\OPtitle[The Bi-Channel Networking Paradigm for Database Systems in the Cloud]{The Bi-Channel Networking Paradigm for\\ Database Systems in the Cloud}



\OPauthor[0009-0003-0169-3478]{Georg Kreuzmayr}
\authornote{Work done while at Technische Universität München}
\email{georg@tigerbeetle.com}
\affiliation{%
  \institution{TigerBeetle}
  \country{}
}

\OPauthor[0000-0001-5295-1316]{Muhammad El-Hindi}
\email{muhammad.el-hindi@tum.de}
\affiliation{%
  \institution{Technische Universität München}
  \country{}
}

\OPauthor[0009-0005-4231-6104]{Benjamin Wagner}
\email{benjamin.wagner@firebolt.io}
\affiliation{%
  \institution{Firebolt Analytics}
  \country{}
}

\OPauthor[0000-0002-1602-4512]{Tobias Ziegler}
\authornotemark[1]
\email{tobias@tigerbeetle.com}
\affiliation{%
  \institution{TigerBeetle}
  \country{}
}

\OPauthor[0000-0001-5676-8017]{Viktor Leis}
\email{leis@in.tum.de}
\affiliation{%
  \institution{Technische Universität München}
  \country{}
}


\renewcommand{\shortauthors}{Kreuzmayr et al.} 


\begin{OPabstract}
When network links were slow, cloud and distributed database systems could rely on generic kernel abstractions and treat network communication as a black box.
With today's fast cloud networks, this approach breaks down: database performance becomes limited by the CPU overhead of the kernel TCP stack.
Replacing TCP with user-space UDP can reduce this overhead, but it requires reimplementing essential guarantees, such as reliability and ordering.
To solve this conundrum, database systems should no longer treat networking as a black box but co-design it with database operations.
We propose the bi-channel paradigm for database systems, which separates communication into two channels:
A high-per\-for\-mance data path for latency- and bandwidth-sensitive operations, and a reliable control path for coordination and recovery.
We implement the paradigm by combining user-space UDP and kernel-based TCP, though other stack combinations are possible.
This design exploits modern NIC capabilities while preserving TCP's reliability.
We demonstrate the paradigm's efficiency and simplicity in two representative settings:
a distributed shuffle saturating 200 Gbit/s with three CPU cores, and a replicated key-value store processing millions of messages per second.
\end{OPabstract}


\maketitle


\section{Introduction}
\label{section:introduction}

\noindent\textbf{Cloud database systems.}
Networking is a major performance concern in cloud database systems.
Online transaction processing (OLTP) systems, such as AWS Aurora~\cite{DBLP:conf/sigmod/VerbitskiGSBGMK17}, require low, predictable network latency to achieve low transaction latency.
Online analytical processing (OLAP) systems such as BigQuery~\cite{DBLP:journals/pvldb/0001GLRSTVADMPS20}, Snowflake~\cite{DBLP:conf/sigmod/DagevilleCZAABC16}, and Firebolt~\cite{DBLP:conf/vldb/PasumanskyW22} rely on parallel execution across many nodes to process large datasets efficiently~\cite{DBLP:journals/pvldb/0001GLRSTVADMPS20, DBLP:conf/sigmod/DagevilleCZAABC16}.

\begin{figure}[t]
    \centering
    \includegraphics[width=\linewidth]{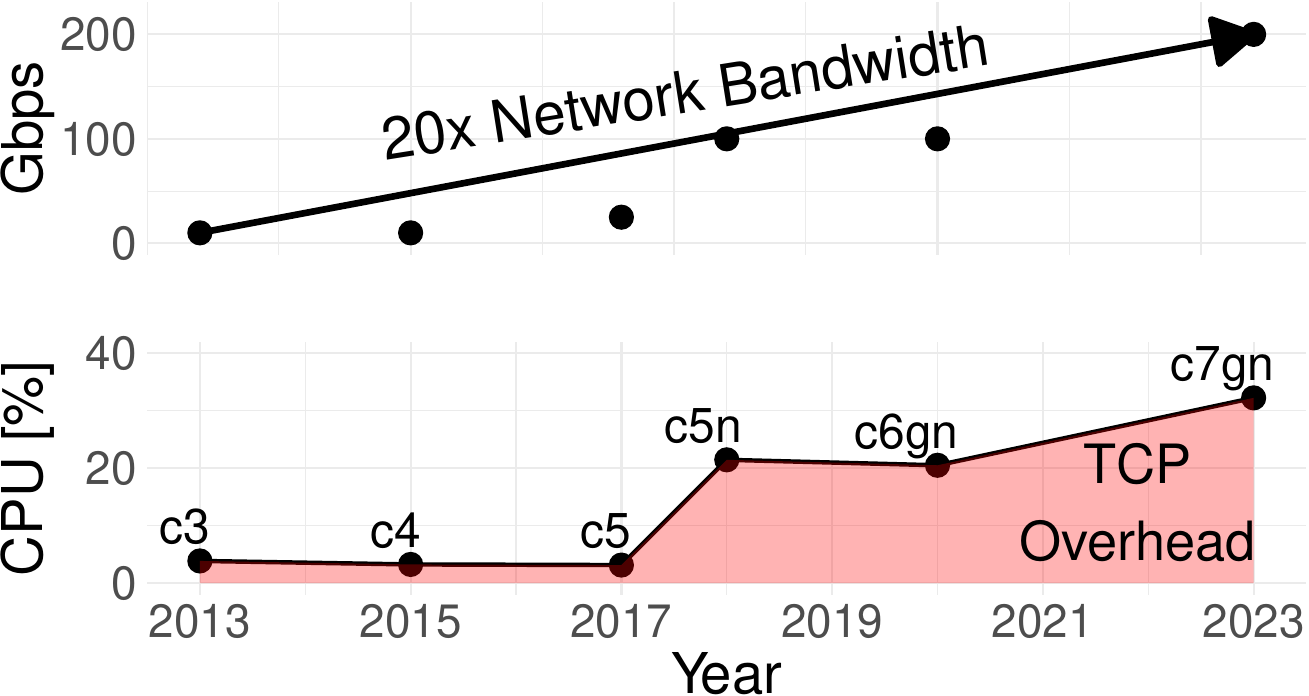}
    \caption{AWS Ethernet bandwidth over the last decade (top). CPU utilization of the kernel TCP stack while saturating full-duplex bandwidth (bottom). Measurements use io\_uring in poll mode with 32 TCP connections and a 256~KiB buffer per connection.}
    \label{fig:network_performance_trends}
\end{figure}

\noindent\textbf{Network hardware is fast.}
Historically, network hardware constrained the performance of distributed data systems~\cite{DBLP:journals/ftdb/BabuH13}.
For example, network bandwidth limited the throughput of data-intensive operators such as distributed joins~\cite{DBLP:journals/pvldb/RodigerMK015}.
However, this has changed with recent advances in network technology: cloud Ethernet networks now offer high bandwidth at low cost.
As \Cref{fig:network_performance_trends} (top) shows, in 2013 the fastest network interface provided 10~Gbps; today, instances such as c7gn provide 200~Gbps, and EC2 instances with 600~Gbps have recently been introduced~\cite{amazon-ec2-600G}.

\noindent\textbf{Networking clogs the CPU.}
While network bandwidth has increased rapidly, CPU performance has improved much more slowly.
\Cref{fig:network_performance_trends} (bottom) shows the CPU utilization required to saturate full-duplex bandwidth on AWS EC2 instances over time.
In 2013, saturating 10~Gbps on a c3 instance required about 3\% CPU.
In contrast, saturating 200~Gbps on a c7gn instance requires over 30\% CPU (20 of 64 cores).

\begin{figure}
    \centering
    \includegraphics[width=\columnwidth]{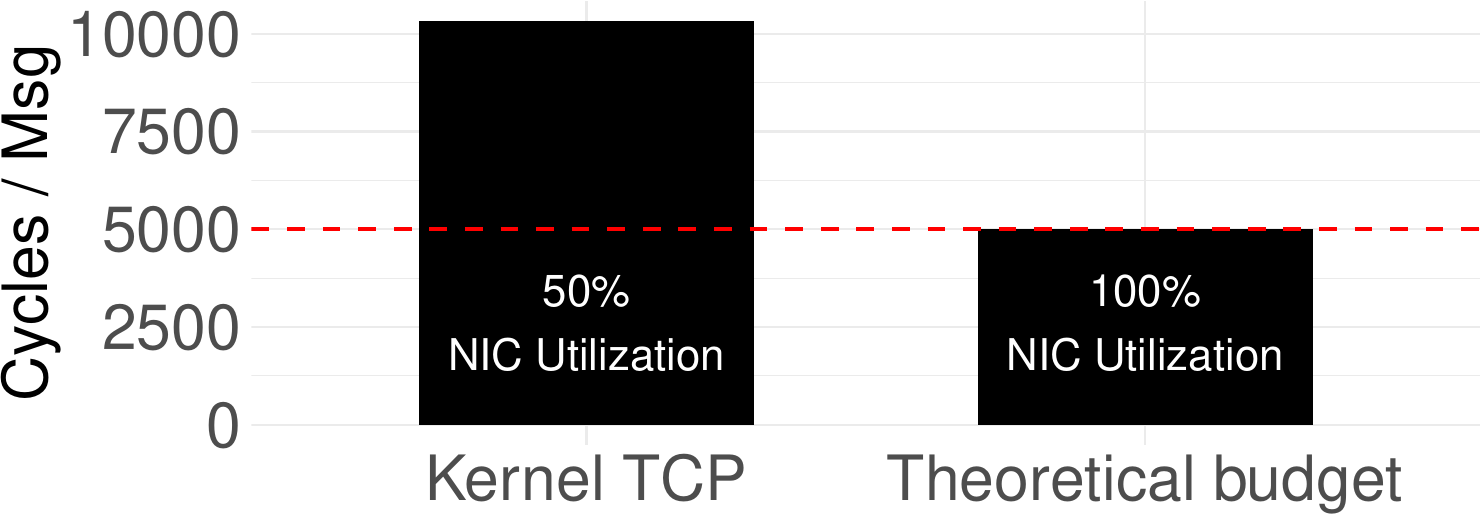}
    \caption{Kernel TCP requires $\approx$\,2$\times$ more cycles/message to fully utilize the NIC (measured with perf while running sockperf~\cite{sockperf} + jumbo frames). A budget of $\approx$\,5k cycles/message is implied by a NIC with 30~M messages/s. The kernel stack is CPU-bound, achieving only $\approx$\,50\% NIC utilization.}
    \label{fig:cycles_intro}
\end{figure}

\noindent\textbf{TCP is inefficient.}
A major contributor to this CPU cost is the kernel TCP stack, which most database systems use for inter-node communication.
TCP provides congestion control, reliability, in-order delivery, segmentation, and stream semantics, but these features impose substantial overhead at high bandwidths (\Cref{fig:network_performance_trends}).
For high-performance OLTP systems, a recent experimental study~\cite{DBLP:conf/cidr/Stonebraker} showed that the CPU overhead of the network stack had become the ``high pole in the tent'', limiting overall transactional throughput.
The reality is even more severe: modern network interface cards (NICs) can handle message rates far beyond what the kernel TCP stack allows~\cite{DBLP:conf/damon/JasnyE0B25}.
For example, the NIC in a c7gn instance can reach roughly 30~M messages/s, which corresponds to about 5{,}000 CPU cycles/message on a 64-core 2.6~GHz CPU ($2.6\times 10^9 \cdot 64 / (30\times 10^6)$).
However, \Cref{fig:cycles_intro} shows that kernel TCP consumes roughly 2$\times$ this budget, leaving the NIC underutilized.

\noindent\textbf{Custom UDP-based protocols are complex.}
Compared to TCP, UDP can be more CPU efficient, and, in principle, database systems could build their networking entirely on top of UDP.
However, doing so requires re-implementing key transport features.
For example, data shuffling for a join needs reliability, which is non-trivial to achieve efficiently on UDP.
Similarly, client connection handling requires congestion control mechanisms to fairly share bandwidth among all clients.
Thus, TCP is costly, but rebuilding its semantics inside the DBMS is also expensive and error-prone.

\begin{figure}
    \centering
    \includegraphics[width=\columnwidth]{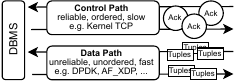}
    \caption{Bi-channel networking uses two coordinated transport paths: a high-throughput, unordered data channel and a reliable, ordered control channel.}
    \label{fig:bi-channel}
\end{figure}

\noindent\textbf{The bi-channel communication paradigm.}
We observe that database operators do not always require the full TCP feature set.
We therefore introduce the \emph{bi-channel paradigm}, an architectural pattern for distributed database systems that decouples the control and data paths at the networking layer.
Instead of committing to a single transport protocol, the system coordinates two channels: (i) a high-throughput, unordered data channel for bulk transfers and (ii) a reliable, ordered control channel for coordination, retransmission, and failure handling (\Cref{fig:bi-channel}).
This design shifts part of the transport boundary into the DBMS, allowing each operator (e.g., shuffle or replication) to choose the channel that matches its required guarantees.
As a result, systems can better exploit modern NICs without re-implementing TCP semantics in user space.
Analogous to how multi-core and SIMD reshaped CPU-bound processing, the bi-channel paradigm reshapes how distributed database systems use high-speed networks.

\noindent\textbf{Contributions.}
We make three contributions:
\begin{enumerate}
\item We propose the \emph{bi-channel paradigm}, a design principle for distributed database systems that separates control and data paths at the transport layer. The paradigm generalizes how a DBMS can combine heterogeneous network stacks to match operator requirements.
\item We derive \emph{vendor-agnostic design guidelines} for high-per\-for\-mance user-space networking in cloud environments. Using AWS as a case study, we analyze constraints such as per-flow bandwidth limits and NIC multi-queue parallelism and distill techniques that generalize across clouds.
\item We demonstrate the paradigm in two database use cases: (i)~a distributed shuffle operator that saturates a 200~Gbps link using three CPU cores and (ii) a replicated key-value store that achieves low latency at high message rates.
\end{enumerate}

\noindent\textbf{Outline.}
The paper is organized as follows.
\Cref{sec:motivation} motivates the problem in more depth and reviews limitations of current networking stacks and the role of kernel bypass.
\Cref{section:co-design} presents the bi-channel paradigm and its design principles.
\Cref{sec:cloud_nic} uses AWS as a representative environment to show how to implement and tune the approach, highlighting pitfalls and distilling practical guidelines.
\Cref{section:shuffle} evaluates using the bi-channel paradigm in a high-bandwidth shuffle operator, and \Cref{section:kvstore} evaluates it in a low-latency replicated key-value store.
We discuss related work in \Cref{section:related_work} and conclude in \Cref{section:conclusion}.


\section{Motivation \& Background}
\label{sec:motivation}

Distributed database systems depend on efficient network communication.
Although prior work has explored new hardware technologies such as RDMA, public-cloud networking is still predominantly Ethernet-based.
Consequently, TCP and UDP remain the only transport protocols that are universally available across cloud platforms.

\noindent\textbf{Kernel-space TCP.}
TCP provides reliable, in-order delivery over an unreliable network.
The kernel handles loss recovery, reordering, segmentation, and flow/congestion control.
Database systems typically use kernel TCP for three reasons.
First, when network bandwidth was limited, TCP overhead was rarely a primary performance concern.
Second, TCP offers strong end-to-end guarantees behind a stable interface, which simplifies DBMS implementations.
Third, kernel TCP is widely deployed and operationally mature.

\begin{figure}
      \centering \includegraphics[width=\columnwidth]{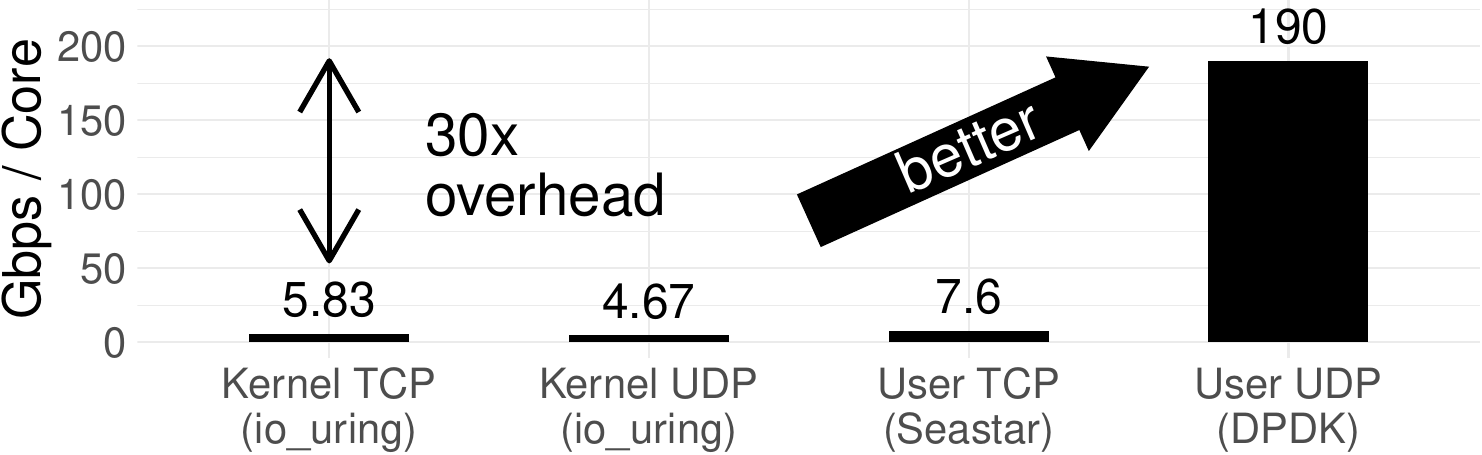}
      \caption{Full-duplex throughput per core between two c7gn.16xlarge instances for kernel- and user-space TCP and UDP, using 256~KiB messages.}
      \label{fig:tcp_perf}
\end{figure}

Today, on high-bandwidth links, enforcing reliability, ordering, and stream semantics can consume a substantial fraction of CPU capacity.
Our benchmarks between two c7gn.16xlarge instances (each with a 200~Gbps NIC), using \texttt{io\_uring} in poll mode~\cite{DBLP:journals/corr/abs-2512-04859}, quantify this cost.
As \Cref{fig:tcp_perf} shows, kernel TCP reaches 150~Gbps per direction while consuming 26 CPU cores (over 40\% of the instance's total CPU resources), i.e., an average of 5.8~Gbps per core.
As network bandwidth continues to increase (e.g., 600~Gbps today and higher rates expected), this imbalance is likely to worsen.

\noindent\textbf{Kernel-space UDP.}
UDP is message-oriented and does not provide reliability or in-order delivery.
Given its simpler semantics, one might expect kernel UDP to be more efficient than kernel TCP.
However, \Cref{fig:tcp_perf} shows that kernel UDP achieves slightly lower throughput in our setup.
The reason is that TCP benefits from throughput-oriented kernel optimizations (e.g., packet batching via Nagle's algorithm), whereas the kernel UDP path is often tuned for low latency and minimal buffering.

\noindent\textbf{User-space TCP.}
To reduce kernel overheads, researchers have developed user-space TCP stacks, where the transport protocol runs entirely in user space.
Examples include mTCP~\cite{DBLP:conf/nsdi/JeongWJJIHP14}, IX~\cite{DBLP:conf/osdi/BelayPKGKB14}, TAS~\cite{DBLP:conf/eurosys/KaufmannSPSKA19}, F-Stack~\cite{fstack}, LUNA~\cite{DBLP:conf/usenix/ZhuSXSFMCWWLYCL23}, and Seastar~\cite{seastar}.
We use Seastar, the stack used by ScyllaDB~\cite{scylla}, in our benchmarks.
Seastar employs a shard-per-core architecture to reduce context switching and lock contention.
Nevertheless, \Cref{fig:tcp_perf} shows only a $\sim$30\% throughput improvement over the \texttt{io\_uring}-based kernel baseline in our configuration.
Prior work (e.g., IO-TCP~\cite{DBLP:conf/nsdi/KimNGKYP23} and LUNA~\cite{DBLP:conf/usenix/ZhuSXSFMCWWLYCL23}) reports similar overhead trends.

\begin{figure}
  \centering \includegraphics[width=\columnwidth]{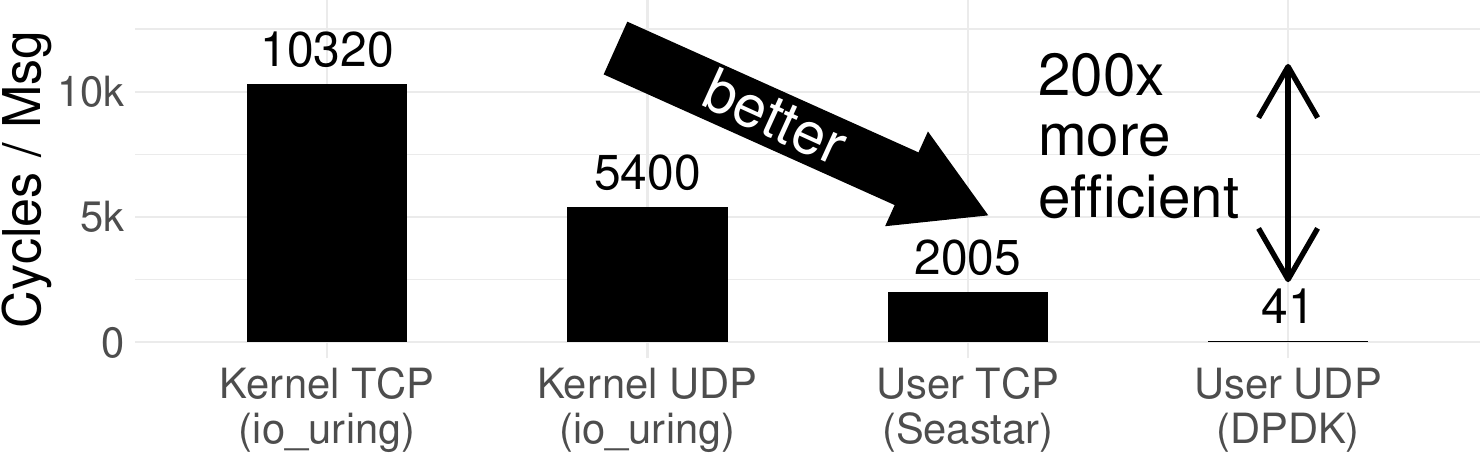}
  \caption{CPU cycles per message between two c7gn.16xlarge instances for kernel- and user-space TCP and UDP, using 64-byte messages.}
  \label{fig:cycles}
\end{figure}

\noindent\textbf{User-space UDP.}
User-space network stacks are often built on the Data Plane Development Kit (DPDK), a collection of libraries and drivers designed for high-performance packet processing in user space.
Unlike TCP, UDP can be implemented directly on DPDK without complex connection state.
In our experiments, a DPDK-based UDP stack reaches 190~Gbps using only \textbf{one} CPU core.
This means DPDK's raw packet processing capability is more than 25 times faster than the kernel- and user-space TCP network stacks.
Message rate is also critical for key-value stores and OLTP workloads.
\Cref{fig:cycles} reports cycles per message for 64-byte messages: user-space UDP reduces CPU cost by orders of magnitude relative to kernel TCP and also substantially outperforms user-space TCP.

However, user-space UDP also comes with practical limitations.
For example, sharing a NIC across multiple processes is challenging without additional multiplexing support.
Moreover, applications that need reliability, ordering, or segmentation must implement these mechanisms above UDP.

\noindent\textbf{Summary.}
Existing options leave an unsatisfying trade-off between strong semantics and efficiency.
\Cref{tab:network_stack_comparison} summarizes the qualitative characteristics of the four stacks.

\begin{table}[h]
  \centering
  \setlength{\arrayrulewidth}{1.5pt}
  \renewcommand{\arraystretch}{1.3}
  \caption{Qualitative comparison of TCP and UDP stacks}
  \label{tab:network_stack_comparison}
  \begin{tabular}{lcccc}
  \hline
  \multicolumn{1}{l}{} & \multicolumn{2}{c}{\textbf{Kernel-space}} & \multicolumn{2}{c}{\textbf{User-space}} \\
   & \textbf{TCP} & \textbf{UDP} & \textbf{TCP} & \textbf{UDP} \\
  \hline
  \textbf{ordering}      & \checkmark & X & \checkmark & X \\
  \textbf{reliability}   & \checkmark & X & \checkmark & X \\
  \textbf{abstraction}   & stream & message & stream & message \\
  \textbf{CPU cost}      & high & high & high & low \\
  \textbf{impl. effort} & -- & -- & high & low \\
  \hline
  \end{tabular}
\end{table}

Kernel TCP provides reliability, in-order delivery, and segmentation, but it can be CPU-intensive and may enforce stronger guarantees than many database operators require.
In particular, reliable in-order delivery can trigger head-of-line blocking~\cite{DBLP:conf/globecom/ScharfK06}: loss or delay of an early packet stalls delivery of later packets even when the application could process messages out of order.
This increases tail latency and becomes more pronounced at high data rates~\cite{DBLP:journals/micro/ShalevABS20}.
While research proposals for alternative transports exist~\cite{DBLP:journals/corr/abs-2210-00714}, their cloud availability and adoption remain unclear.
Conversely, user-space UDP achieves much higher efficiency but lacks reliability and requires applications to implement features such as segmentation for messages larger than an MTU.

However, we observe that the transport guarantees required by database operators often fall between these extremes.
For example, a join/shuffle operation needs eventual delivery of all tuples, but not in-order delivery.
This motivates our main research question: can we combine TCP-like guarantees with the efficiency of user-space UDP, and tailor transport semantics to the needs of individual database operators?


\section{The Bi-Channel Paradigm for Database Systems}
\label{section:co-design}
To leverage the strengths of different network stacks, we introduce the \emph{bi-channel paradigm}, which enables distributed systems to benefit from both high-performance user-space UDP and the robustness of kernel-space TCP.

\begin{figure}
    \centering
    \includegraphics[width=0.8\columnwidth]{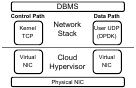}
    \caption{Example bi-channel network stack for a cloud DBMS.}
    \label{fig:network_stacks}
\end{figure}

\noindent\textbf{Separating control and data communication.}
Database networking naturally decomposes into control and data communication.
Control messages (e.g., transaction coordination and query planning) typically require reliability and in-order delivery, whereas data transfers (e.g., tuple shuffling) primarily benefit from low CPU overhead and high throughput.
As shown in \Cref{fig:network_stacks}, a cloud DBMS can therefore use kernel TCP for the control channel and a DPDK-based user-space UDP stack~\cite{dpdk} for the data channel.

\noindent\textbf{Kernel TCP control channel.}
DBMSs exchange control messages across workloads and operators.
In OLTP, nodes communicate to execute distributed commit protocols such as two-phase commit~\cite{DBLP:journals/tods/MohanLO86}.
In OLAP, control traffic coordinates data exchange and scheduling decisions (e.g., partition assignment and load balancing).
Because control traffic requires reliability and ordering guarantees, kernel-based TCP remains the most suitable choice for the control path.

\noindent\textbf{User-space UDP data channel.}
The data channel targets per\-for\-mance-critical transfers.
Its requirements vary by workload: analytical shuffles emphasize high throughput (\Cref{section:shuffle}), while replication and log shipping in transactional systems emphasize high message rates and low latency (\Cref{section:kvstore}).
User-space UDP with DPDK provides the packet I/O efficiency needed for these hot paths.

\noindent\textbf{DBMS integration.}
Realizing the bi-channel benefits requires co-design between the network path and DBMS components (e.g., physical operators).
This is aligned with a long-standing database practice: bypassing generic OS abstractions on the critical path when they impose unnecessary overhead.
For example, many DBMSs implement custom buffer managers instead of relying on the OS page cache~\cite{DBLP:journals/pacmmod/LeisA0L023}.
Analogously, exposing transport choices to operators can significantly improve networking efficiency.
At the same time, the paradigm avoids an all-or-nothing shift to user-space networking: the kernel control channel provides mature reliability and ordering where needed, reducing the need to re-implement full TCP semantics in the DBMS.

\noindent\textbf{Leveraging virtualized cloud NICs.}
A practical obstacle to combining kernel and user-space stacks on the same interface is that frameworks such as DPDK typically require exclusive access to a NIC.
All major cloud providers mitigate this constraint by allowing multiple virtualized network interfaces (vNICs) to be attached to a single instance with minimal effort~\cite{amazon-multi-nic}.
This enables the architecture shown in \Cref{fig:network_stacks}, where two vNICs are used: one is managed by the kernel (control channel) and the other is bound to DPDK in user space (data channel).

\noindent\textbf{Reliability of cloud Ethernet networks.}
The bi-channel approach relies on the assumption that packet loss is low enough that recovery traffic does not dominate the data channel.
To assess this in practice, we measured packet loss in an AWS data center using a full-duplex UDP ping-pong benchmark between two c7gn.16xlarge instances in the same data center.
We ran the benchmark for one minute at full bandwidth and recorded sender-side loss with a transmission depth of 1024.
As shown in \Cref{tab:packet_loss}, four out of five runs delivered all packets successfully at an average bandwidth of 190~Gbps.
In the remaining run, 19 packets were lost out of 150 million, corresponding to an overall loss rate of 1.3$\times 10^{-7}$ (0.000013\%).
In an additional one-hour run, we observed a single lost packet.
Consistent with these measurements, Google Cloud's performance reporting indicates low packet loss rates (below 0.05\%) across regions~\cite{gcppacketloss}.
Together, these results suggest that cloud Ethernet can support a hot user-space UDP data channel in practice, with infrequent loss recovery.

\begin{table}
    \centering
    \caption{Packet loss for full-bandwidth UDP ping-pong communication between two \texttt{c7gn.16xlarge} instances. In total, 150 million packets were transmitted.}
    \label{tab:packet_loss}
    \begin{tabular}{lccccc}
        \toprule
        \textbf{Run} & 0 & 1 & 2 & 3 & 4 \\
        \midrule
        \textbf{Lost packets} & 0 & 0 & 0 & 19 & 0 \\
        \bottomrule
    \end{tabular}
\end{table}


\section{Case Study: Bi-Channel Paradigm in the Cloud (AWS Example)}
\label{sec:cloud_nic}

The bi-channel paradigm is a general architectural framework, independent of any specific hardware or cloud vendor.
However, the concrete implementation effort and tuning parameters may differ because each platform exposes different network abstractions and performance limits.
Among public clouds, these limits vary substantially: for instance, Google Cloud primarily restricts external egress bandwidth~\cite{gcpbw}, while AWS enforces per-flow limits even for intra-VPC traffic~\cite{amazon-ec2-single-flow}.
Therefore, while the bi-channel principles remain unchanged, realizing a high-performance data path requires adapting to the specific flow-control and queueing semantics of the underlying infrastructure.

\noindent\textbf{AWS as a stress-test environment.}
We use AWS as a representative case study because it imposes the most restrictive intra-cloud networking constraints.
These constraints make it an ideal stress test for the bi-channel design, which must remain efficient even under stringent per-flow and virtualization limits.
The lessons learned from AWS extend to other clouds: the same structured approach and guidelines discussed below can be applied based on the exposed constraints in other clouds.
For example, our experiments use the Elastic Network Adapter (ENA) on \texttt{c7gn.16xlarge} instances, but the derived principles are portable across NIC implementations supporting parallel transmit/receive queues, RSS-based load distribution, and user-space packet I/O.
We next present a mental model of cloud NICs and derive guidelines for building robust, high-throughput data paths using user-space UDP.
All experiments were conducted on Ubuntu 24.04 with DPDK 23.11.2.

\subsection{Mental Model for Cloud NICs}\label{subsec:mental_model}
\begin{figure}
    \centering
    \includegraphics[width=\columnwidth]{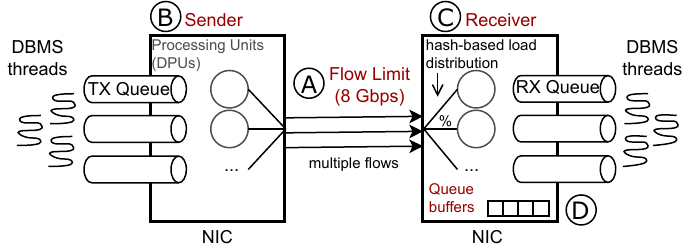}
    \caption{Mental model for cloud NICs: DBMS threads can directly interact with the NIC's hardware queues and use zero-copy communication. However, several pitfalls (A-D) exist that developers must be aware of.}
    \label{fig:nic_model}
\end{figure}

\noindent\textbf{The way of the packet.}
Leveraging user-space frameworks, such as DPDK, with the bi-channel paradigm enables applications to operate very close to the metal -- that is, directly at the NIC.
As shown in \Cref{fig:nic_model}, DBMS threads can directly interact with the NIC's hardware queues.
In particular, when sending packets, an application places them in the transmit (TX) queue (see \Cref{fig:nic_model} left).
Since NICs are highly parallel devices, they have multiple data processing units (DPUs) (circles) that consume packets from the TX queue and send them over the cloud network.
On the receiving NIC (\Cref{fig:nic_model}, right), incoming packets are routed to different DPUs for parallel processing (using a hash-based load distribution).
The DPUs copy the packets via Direct Memory Access (DMA) into receive or queue buffers, after which the DPU places a completion event into the receive (RX) queue.
Application threads at the receiver poll the RX queues and directly read packets from these buffers. 
This mechanism is known as zero-copy communication, where packets are copied only once, directly from the NIC into user-accessible receive buffers in RAM, and are never copied again within RAM.

\noindent\textbf{Potential networking bottlenecks.}
Despite the conceptually simple design shown in \Cref{fig:nic_model}, many pitfalls exist when trying to achieve robust, high-performance networking.
For example, cloud networks have complex topologies~\cite{DBLP:conf/sigcomm/SinghOAAABBDFGK15} and multiple tenants share their underlying infrastructure, forcing cloud providers to impose so-called flow limits on a communication path (\numcircledmod{A}), i.e., a path between two endpoints.
Moreover, slight misconfigurations can lead to bad performance on the sender or receiver side (\numcircledmod{B} \& \numcircledmod{C}), particularly regarding queue buffers \numcircledmod{D} that are the backing memory for entries in the hardware queues.

In the following, we detail pitfalls and derive guidelines for achieving high-performance and robust networking on the data path, when optimizing either for bandwidth (\Cref{subsec:max_bandwidth}) or packet rate and latency (\Cref{subsec:max_packet_rate}).

\subsection{Optimizing for Bandwidth}
\label{subsec:max_bandwidth}
The bi-channel paradigm's data path represents the hot path for database operations.
For example, it is important to optimize this hot path for high-bandwidth data processing in the context of OLAP databases.
In the following sections, we highlight pitfalls and demonstrate how to achieve high bandwidth using user-space networking in the cloud.

\noindent\textbf{Single-flow bandwidth limits.}
Although c7gn.16xlarge instances in AWS offer a 200 Gbps network interface, and using DPDK out-of-the-box in the data path to send data between two instances will surprisingly hit a bottleneck at around 8 Gbps.
This is because AWS enforces a single-flow bandwidth limit restricting the throughput of TCP and UDP connections~\cite{amazon-ec2-single-flow}.
Within one cluster placement group, which we used for our experiments, the limit is 10 Gbps.
Our empirical measurements showed that the throughput is limited to 8 Gbps when using a single UDP port combination.
To achieve higher throughput, it is required to use multiple paths between the communicating endpoints by varying the source and destination ports in the UDP headers\footnote{In AWS, a single-flow is considered a unique 5-tuple TCP or UDP flow defined by source IP and port, destination IP and port, and the next protocol~\cite{amazon-ec2-single-flow}.}.

\begin{figure}
    \centering
    \includegraphics[width=\columnwidth]{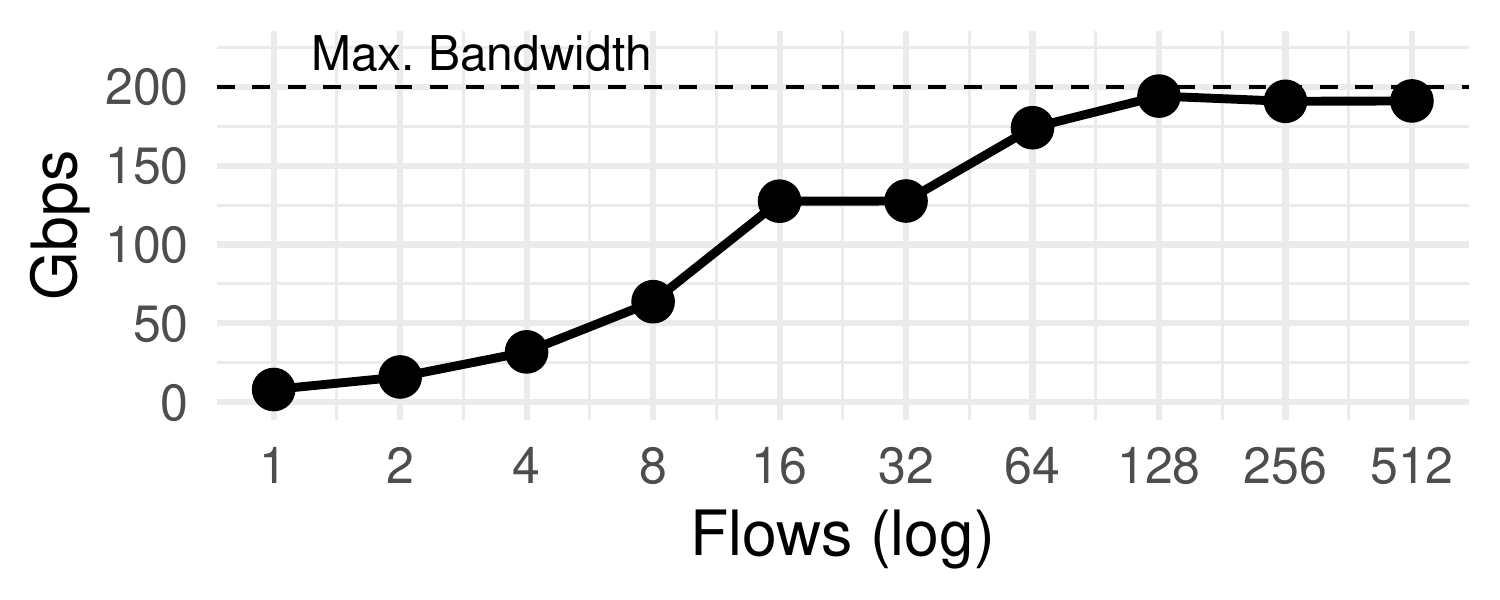}
    \caption{Effect of an increasing number of UDP port combinations (flows) on bandwidth with 8192-byte packet size.}
    \label{fig:scale_ports_bw}
  \end{figure}

\noindent\textbf{Port scaling.}
\Cref{fig:scale_ports_bw} shows the throughput of user-space UDP with an 8192-byte packet size and increasing port combinations, i.e., flows.
We increased the number of port combinations by configuring the sender and receiver threads to use varying source and destination ports.  
Initially, the throughput increases linearly to 16 Gbps using two ports and to 32 Gbps using four ports.
After reaching 130 Gbps with 16 ports, the throughput starts to stagnate.
Only when further increasing the number of ports to 128, we get close to the maximum throughput of EC2's 'c7gn.16xlarge instance type.

\vspace{1ex}
\begin{guideline}[label=rule1]
\noindent\textbf{Guideline 1:}
Beware of bandwidth flow limits. Use multiple flows by varying source and destination ports to increase throughput.
\end{guideline}
\vspace{1ex}

\subsection{Maximizing Packet Rate}
\label{subsec:max_packet_rate}

\noindent\textbf{Single-threaded packet rate.}
In the previous experiment, we used large packet sizes of 8192 bytes.
Hence, data transmission was bottlenecked by the single-flow bandwidth limit, resulting in a packet rate of around 100'000 packets per second.
Repeating the same benchmark with small 64-byte packets stalls at around 2.4 M packets per second.
This time, throughput is well under 8 Gbps, meaning that the limiting factor is not the single-flow bandwidth limit.
To find the root cause of the bottleneck, we analyze whether CPU performance is the limiting factor.

\noindent\textbf{Multi-threaded packet rate.}
Instead of a single thread, we now use multiple threads to send UDP packets and measure the packet rate for half-duplex communication between two servers.
We measure the scalability of the sender's packet rate with the number of threads for a single flow (i.e., UDP port combination):

\begin{center}
    \centering
    \includegraphics[width=0.8\columnwidth]{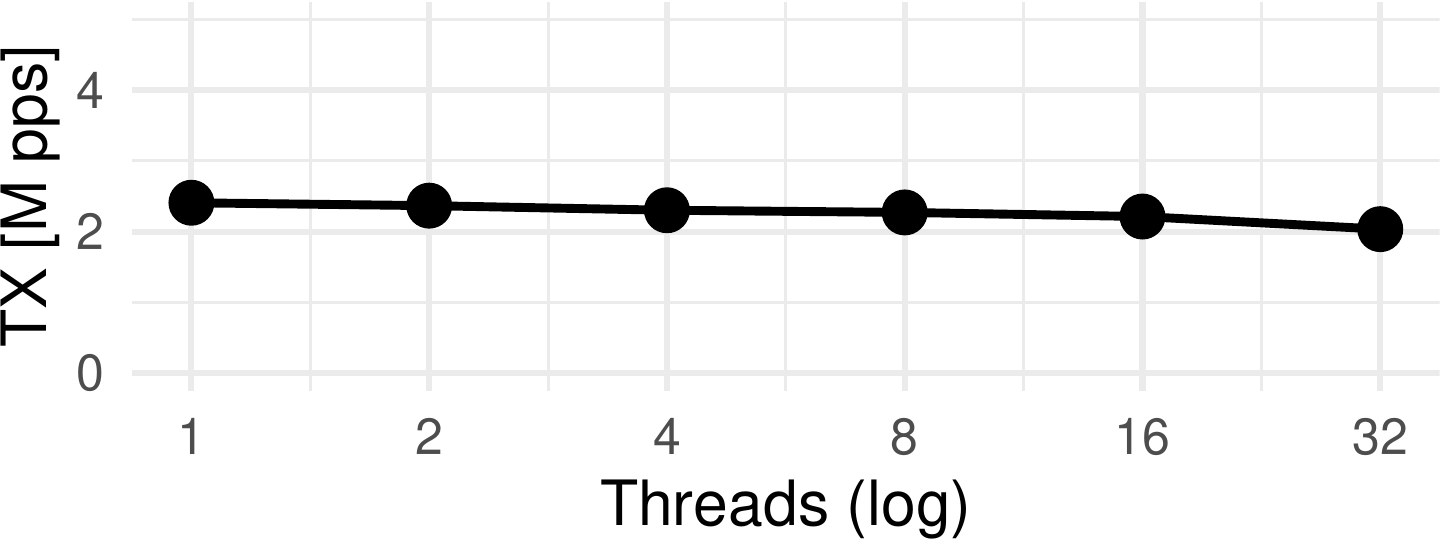}
\end{center}

The results show that simply increasing the thread count while using a single flow achieves the same packet rate as with a single thread (2.4 M packets per second).
Hence, neither the flow limit nor the CPU performance is the limiting factor.

\noindent\textbf{Parallel processing in cloud NICs.}
To understand the root cause of the packet-rate bottleneck, we need to return to our mental model of cloud NICs from \Cref{subsec:mental_model}.
NICs are highly parallel devices that internally utilize multiple DPUs (circles in \Cref{fig:nic_model}), similar to threads, to send packets in parallel.
Similar to sharding in databases, the parallelism is achieved by partitioning work across DPUs using a hash function based on the UDP headers (function $h()$ in \Cref{fig:nic_model}). 
Despite scaling the number of CPU threads in the earlier experiment, all packets were sent as part of a single flow, i.e., using the same source and destination ports, and thus processed by a single DPU.
This deterministic partitioning is important when processing dependent packets, such as in TCP streams, but it can result in load imbalances, which is the root cause of the stagnating packet rate at 2.4 M packets.

\begin{figure}[t]
  \centering
  \begin{subfigure}[b]{0.49\columnwidth}
      \centering
      \includegraphics[width=\textwidth]{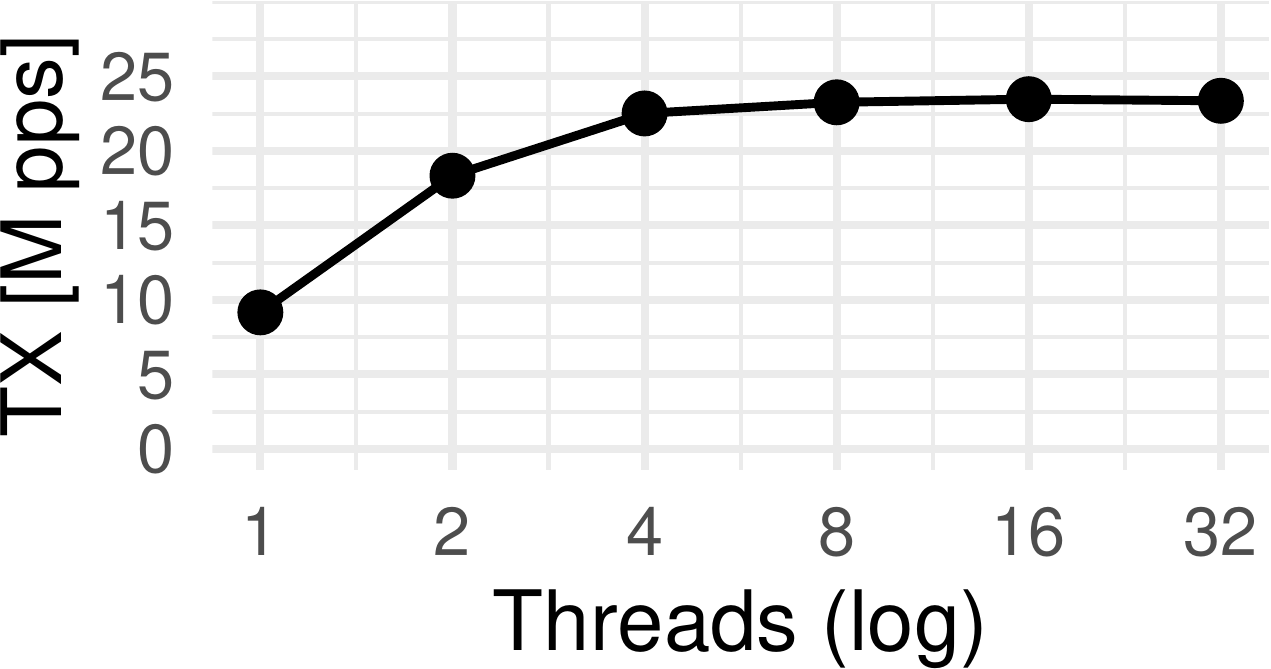}
      \caption{Sender side}
      \label{fig:transmitter_threads}
  \end{subfigure}
  \hfill
  \begin{subfigure}[b]{0.49\columnwidth}
      \centering
      \includegraphics[width=\textwidth]{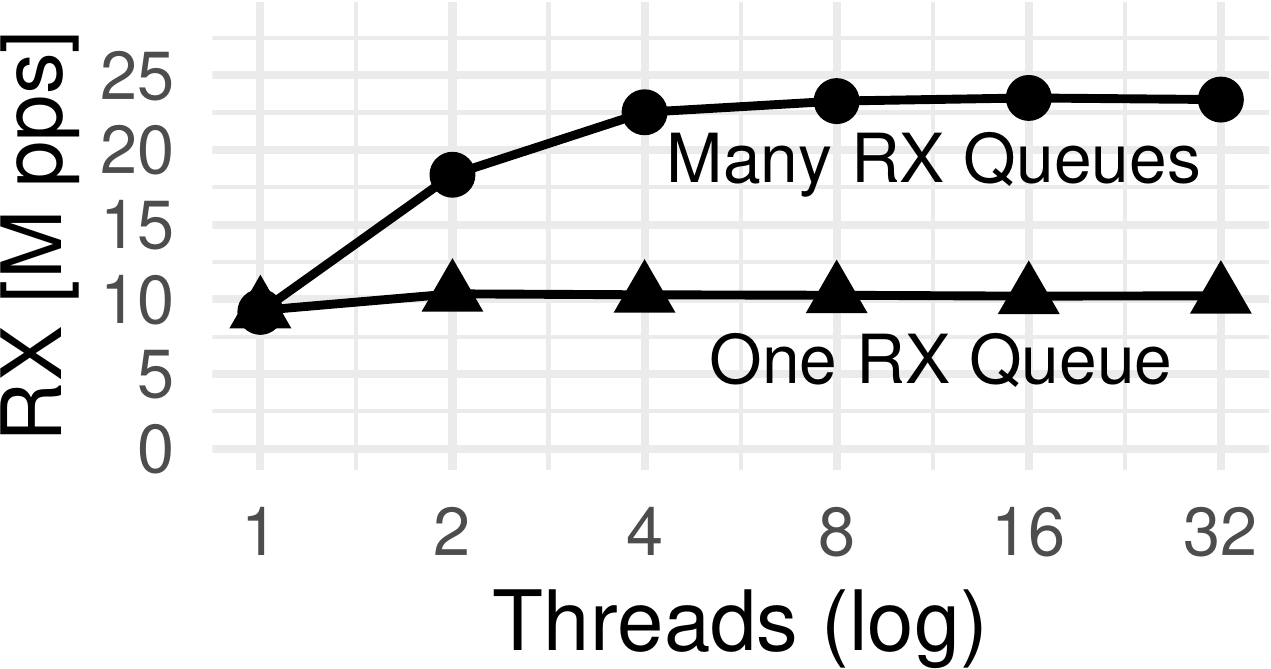}
      \caption{Receiver side}
      \label{fig:recevier_threads}
  \end{subfigure}
  \caption{Thread scalability of sender and receiver side with 64-byte packets between two c7gn.16xlarge instances.}
  \label{fig:thread scaling}
\end{figure}

\noindent\textbf{Sender scaling.}
To overcome this limitation, we apply the port-scaling optimization to use multiple flows and distribute the work across the DPUs.
This figure shows the scalability of the sender's packet rate with the number of UDP ports (flows):

\begin{center}
    \includegraphics[width=0.8\columnwidth]{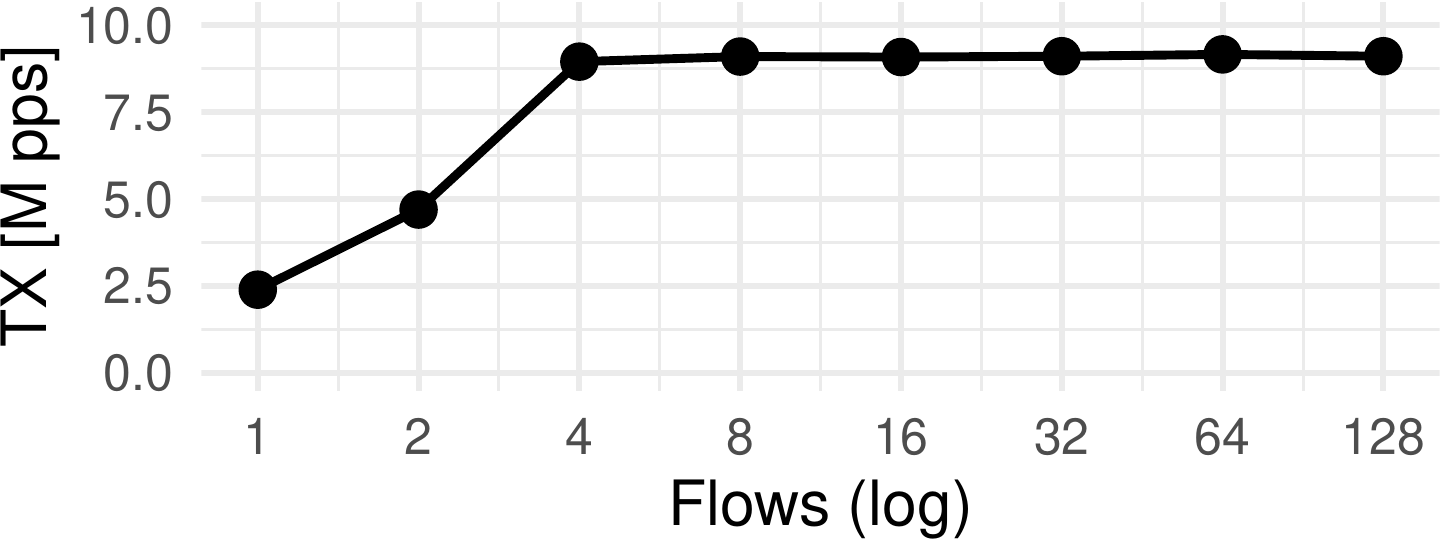}
\end{center}

It demonstrates that using multiple UDP port combinations improves the packet rate to around 9 M packets per second with a single thread.
This confirms that the previously observed rate of 2.4 M was not limited by CPU performance.
In practice, it is advisable to perform micro-benchmarks to determine how many port combinations are required to maximize the packet rate.

\vspace{1ex}
\begin{guideline}[label=rule2]
\noindent\textbf{Guideline 2:}
Beware of packet rate flow limits. Use multiple flows to achieve high packet rates.
\end{guideline}
\vspace{1ex}

\noindent\textbf{Thread parallelism and port scaling.}
Having determined the single-threaded packet rate bottleneck, we now attempt to increase the overall packet rate by employing thread parallelism again.
We use 64 different UDP port combinations while increasing the number of threads used on the sender side.
As shown in \Cref{fig:transmitter_threads}, this configuration can scale the transmitted packets per second up to around 25 M.
However, \Cref{fig:recevier_threads} (triangle line) shows that with the same configuration (i.e., using 64 different flows), the receiver side does not scale in the same way.

\noindent\textbf{Receiver scaling.}
Examining the receiver side closely, reveals a similar problem as on the sender side:
Packet processing on the receiver is skewed, and not all DPUs are processing packets.
The triangle configuration in \Cref{fig:recevier_threads} used UDP port combinations that caused all packets to be distributed to the same DPU, limiting the receiver side's packet rate.
This was configured deliberately and shows the worst-case effect of skewed packet distribution.

\noindent\textbf{Queue targeting.}
A NIC typically exposes multiple RX queues.
Polling all RX queues from every thread, however, would be inefficient.
Instead, it is preferable to determine the RX queue to which each packet is steered, so that the corresponding thread can process it directly.
This capability enables efficient implementations, such as sharded key-value stores, in which requests are routed to the worker thread responsible for the target shard.
The mechanism that enables such packet steering is commonly known as receive side scaling (RSS) and is widely supported by modern physical and virtual NICs.
Because RSS is deterministic, a given packet is always assigned to the same RX queue.
AWS, Azure, and GCP document the hash functions used by their RSS implementations~\cite{ena-github-dpdk,gcp-github-dpdk,azure-github-dpdk}.
This allows the sender to precompute the receiving thread.
We refer to this technique as \textit{Queue Targeting}.
It enables applications to control packet routing in user space and balance load across cloud instances.

\noindent\textbf{The effect of queue targeting.}
The circle configuration in \Cref{fig:recevier_threads} uses 64 different UDP port combinations per thread while leveraging our queue targeting technique.
This way, we achieve scalability not only on the sender side but also on the receiver side.

\vspace{1ex}
\begin{guideline}[label=rule3]
\noindent\textbf{Guideline 3:}
Leverage queue targeting to control packet routing from the application level and avoid RSS bottlenecks.
\end{guideline}
\vspace{1ex}

\begin{figure}[t]
    \centering
    \begin{subfigure}[b]{0.49\columnwidth}
        \includegraphics[width=\textwidth]{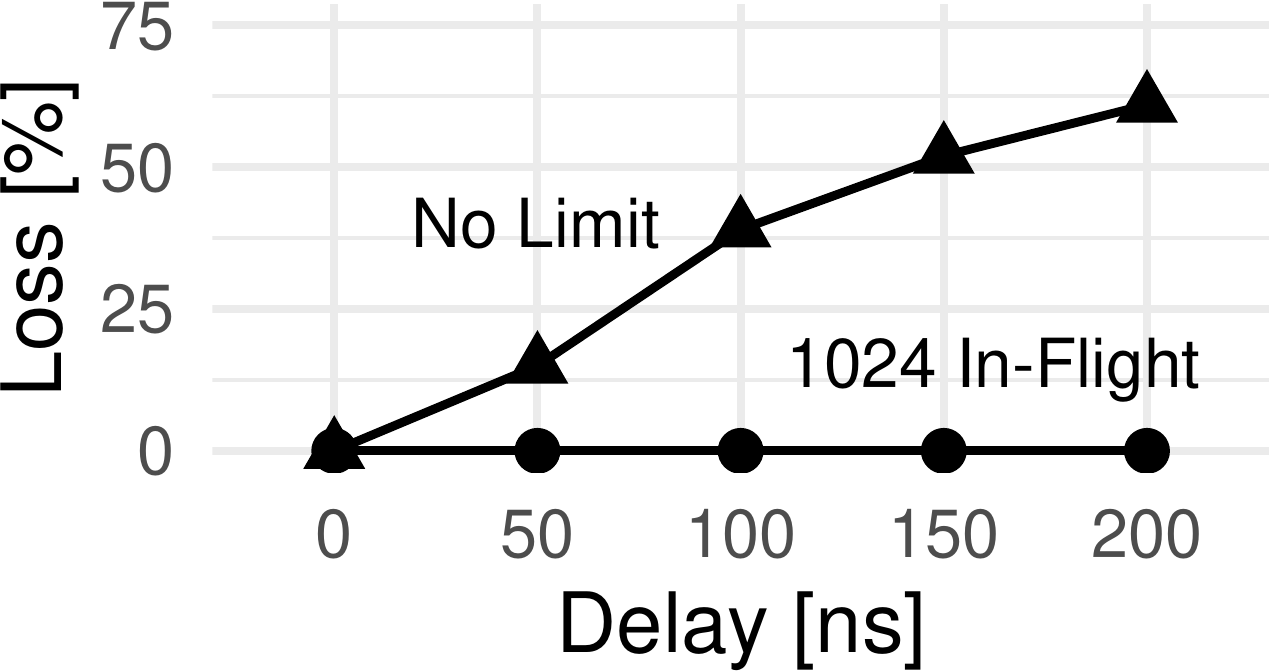}
        \caption{In-flight limit configuration.}
        \label{fig:packet_loss_no_vs_feedback}
    \end{subfigure}
    \hfill
    \begin{subfigure}[b]{0.49\columnwidth}
        \includegraphics[width=\textwidth]{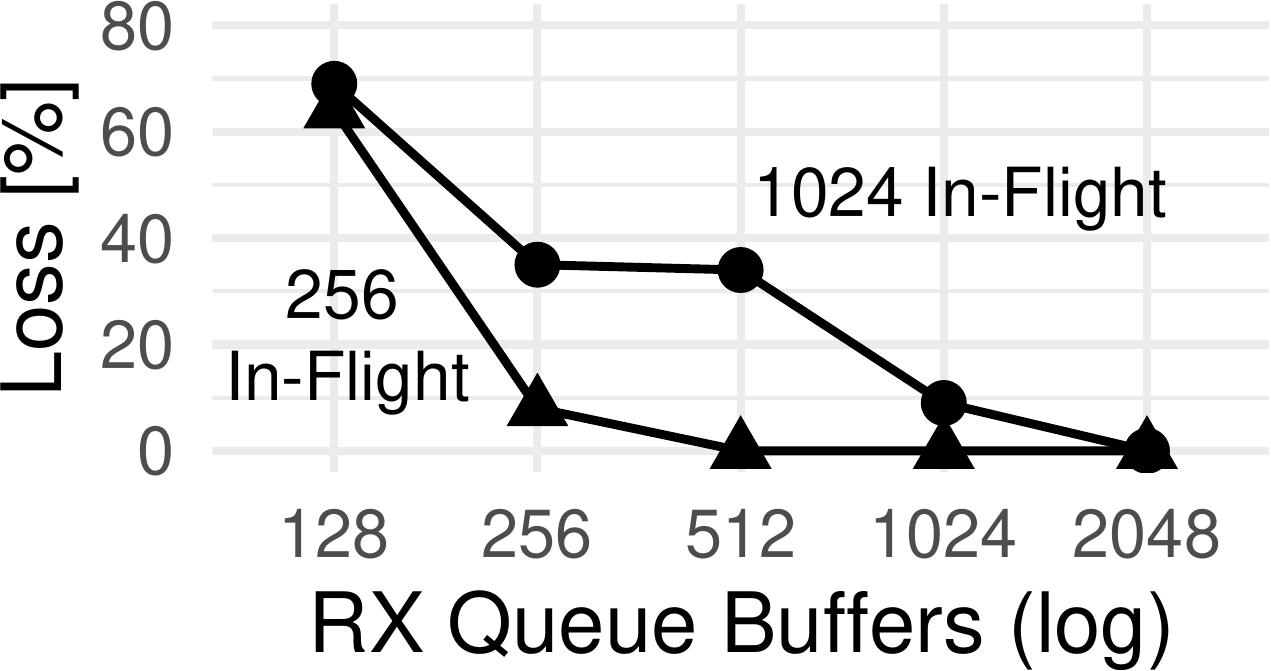}
        \caption{RX descriptor configurations.}
        \label{fig:packet_loss_rx_desc}
    \end{subfigure}
    \caption{Packet loss for different in-flight and RX queue configurations.}
\end{figure}

\subsection{Minimizing Packet Loss}
As with performance, network reliability for user-space UDP in the cloud does not come out of the box.
Yet, reliability is critical for applications to benefit from a high-performance data path.

\noindent\textbf{Packet loss in low load scenarios.}
Under low load, applications rarely observes packet loss when using user-space UDP.
For example, in the experiment for \Cref{tab:packet_loss}, we limited the number of in-flight ping-pong packets to only 1024.

\noindent\textbf{Packet loss in high load scenarios.}
However, the situation is different under high load when applications rely on user-space UDP. 
To illustrate the problem, \Cref{fig:packet_loss_no_vs_feedback} shows the packet loss for single-threaded communication using 64-byte packets for two settings.
In the "No-Limit" (triangle line) setting, we do not limit the number of outstanding requests, which increases receiver load over time.
Due to the lack of a control flow mechanism in user-space UDP, the packet loss rate begins to skyrocket.

\noindent\textbf{Application-side load-balancing.}
The bi-channel paradigm addresses this problem by enabling application-specific control flow.
This is illustrated with the "1024 In-Flight" setting in \Cref{fig:packet_loss_no_vs_feedback}.
In this setting, the receiving server can use the kernel-based control path to report to the sender which packets arrived successfully.
This way, the sender can avoid overloading the receiver by sending more than 1024 packets.
As shown in \Cref{fig:packet_loss_no_vs_feedback} (circle line), this application-side load balancing mechanism avoids increasing packet loss even when the receiver is busy.

\noindent\textbf{Queue buffers.}
A key question when implementing application-side load-balancing is how many in-flight requests to allow.
In other words, how to set the in-flight packet rate for an application or workload?
To answer this question, we need to understand the inner workings of an NIC in depth.
In particular, developers must be aware of the number of queue buffers (\numcircledmod{D} in \Cref{fig:nic_model}).
Queue buffers serve as the backing store for incoming network requests.
That is, when a DPU places a packet into the RX queue for the application to consume, it must allocate a queue buffer.
However, when the queue buffers are exhausted, the NIC will simply drop the incoming UDP packet.
The number of buffers is limited and configurable during device startup.
For example, the c7gn.16xlarge instance type supports between 128 and 8192 queue buffers.

\noindent\textbf{Choosing an in-flight packet rate limit.}
A common assumption is that if the number of in-flight packets is kept below the number of RX queue buffers, the NIC should have enough capacity to receive all packets without drops.
However, our experiments show that this assumption does not always hold and that the in-flight limit must be chosen with care.

\Cref{fig:packet_loss_rx_desc} shows the packet loss for a varying number of RX queue buffers and two different in-flight limits.
To simulate a busy receiver, we perform 100 ns of work in a hot loop per received packet, as described in \Cref{fig:packet_loss_no_vs_feedback}.
When the in-flight limit is higher than the number of RX queue buffers, the RX queue (\Cref{fig:nic_model} right) runs out of buffers, and the packet loss rate goes up.

Even when the in-flight limit equals the number of RX queue buffers, we observe a packet loss rate of around 10\%.
This happens because not all RX descriptors are available for receiving packets at any given time.
With a limit of 256 packets, the loss is almost eliminated once the RX queue has 512 or more buffers.
With 1024 in-flight packets, packet loss disappears with 2048 buffers.

\vspace{1ex}
\begin{guideline}[label=rule4]
   \noindent\textbf{Guideline 4:}
   Use application-side load-balancing via the control path to limit the number of in-flight packets.
   As a rule of thumb, choose the in-flight packet rate relative to the RX queue buffers.
\end{guideline}
\vspace{1ex}

While our case study focuses on AWS, the mental model and best practices derived here, such as Guideline~4, capture vendor-independent principles that can be applied to other cloud platforms.
We next show how the bi-channel paradigm integrates with database systems and can be applied in concrete database use cases.


\section{Use Case 1: Efficient Data Shuffling in Distributed Joins}
\label{section:shuffle}
This section demonstrates how the bi-channel paradigm and low-level NIC optimizations presented previously are applied in database-specific use cases.
We focus specifically on distributed joins due to their intensive network communication requirements.
Modern analytical query engines process queries involving petabytes of data \cite{DBLP:conf/nsdi/VuppalapatiMATM20, DBLP:journals/pvldb/RenenHPVDNLSKK24}, distributing and partitioning data across multiple nodes.
During a join, data is frequently repartitioned -- commonly referred to as "shuffling" -- across the network based on join keys.
This shuffle operator efficiently repartitions data across nodes via the network, making it a good fit for illustrating the benefits of the bi-channel paradigm.

\subsection{Anatomy}
\label{subsec:shuffle}

\noindent\textbf{Shuffle operator characteristics.} 
Implementations of the shuffle operator differ widely in both design and execution.
Systems such as Spark\cite{DBLP:conf/hotcloud/ZahariaCFSS10}, or BigQuery~\cite{DBLP:journals/pvldb/0001GLRSTVADMPS20} rely on distributed storage layer for cross-node data transfers .
Recently, systems without dedicated shuffle services have become more prevalent, often employing direct point-to-point communication.
In these modern architectures, each node can directly transfer data to every other node \cite{DBLP:conf/vldb/PasumanskyW22}, i.e., all-to-all.
Joins and aggregations typically employ hash-based partitioning functions to identify the destination node for each data item.
In this case study, we focus specifically on a point-to-point hash-based shuffle implementation.

\noindent\textbf{Communication characteristics.}
Before applying the bi-channel paradigm, it is important to understand the shuffle operator's communication characteristics.
We identify three main traits:
First, the primary goal of any shuffle implementation is to achieve high bandwidth, as data transfer time heavily influences overall query execution time.
Second, since we focus on hash-based shuffle operations, the order of tuple arrival is not important.
Third, reliability is crucial -- all data must eventually be processed without loss.
Mapping these characteristics into the bi-channel paradigm, we see that the first two align well with the user-space UDP data path.
As demonstrated in \Cref{sec:motivation}, a single core reaches nearly 200 Gbps on the data path, making it an optimal choice, especially given the shuffle's insensitivity to tuple ordering.
The third attribute -- reliability -- necessitates a reliable communication mechanism to guarantee complete data delivery.
This divergence in requirements is precisely what the bi-channel paradigm addresses, making it a natural and effective fit.

\subsection{Bi-Channel Shuffle}
\label{subsec:bi-channel_shuffle}
This section demonstrates how to use the bi-channel paradigm in a shuffle operator.
Since high throughput is crucial for the shuffle, we first leverage the fast user-space channel to meet this performance objective.
We then discuss the qualitative advantages of the secondary, slower channel using kernel-space TCP.

  \begin{figure}
    \centering
    \includegraphics[width=0.99\columnwidth]{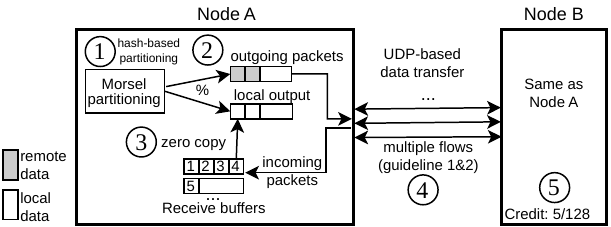}
    \caption{The bi-channel paradigm applied for data shuffling: We use morsel-driven parallelism and a UDP-based data transfer with zero-copy and multiple flows.}
    \label{fig:shuffle_overview}
  \end{figure}

\noindent\textbf{Implementation overview.}
\Cref{fig:shuffle_overview} provides a high-level over\-view of our solution.
It depicts two example nodes that both send and receive data.
The input relation is processed using morsel-driven parallelism~\cite{DBLP:conf/sigmod/LeisBK014}.
Worker threads partition the data into local output buffers for local processing and into outgoing packets for remote nodes \numcircledmod{1}.
Remote communication is handled via the fast path, using UDP packets that are filled incrementally with tuples and sent once full \numcircledmod{2}.
Because each worker also receives data asynchronously from remote nodes, it periodically polls the receive queues for incoming packets stored in receive buffers.
When data arrives, we place pointers to the receive buffers directly into the local output buffer, thereby avoiding unnecessary data copying \numcircledmod{3}.

\noindent\textbf{Thread scalability.}
To ensure that our operator scales with the number of threads, avoiding the NIC from becoming a scalability bottleneck is important.
To achieve this, we apply the principles outlined in the previous chapter: (1) avoiding the bandwidth and packet rate limitations by utilizing multiple parallel flows (guideline~\ref{rule1}+\ref{rule2}) and (2) carefully choosing sender and receiver ports to enable queue targeting (guideline~\ref{rule3}) \numcircledmod{4}.
We encapsulate this port-selection logic in a dedicated component, PortPicker, which is used during cluster provisioning to establish the necessary connections (flows).
At system startup, we configure communication and allocate multiple flows per thread to ensure that single-flow limitations do not constrain system bandwidth. 
This approach means that if we use a single core, it uses multiple parallel flows, effectively multiplexing communication, eliminating the flow bottleneck, and fully parallelizing the NIC.
Once these implications are clearly understood, satisfying them becomes straightforward, enabling our implementation to scale nearly perfectly, as demonstrated by the later experimental evaluation.

\noindent\textbf{Receive-buffer management.}
One crucial aspect we have not yet addressed is the management of queue buffers (i.e., guideline~\ref{rule4}).
Exhausting queue buffers results in significant packet loss, making proper management essential for reliability.
One straightforward solution is to have the remote node synchronously acknowledge each received UDP packet.
However, limiting communication to only a single in-flight packet severely restricts bandwidth utilization.
To achieve higher efficiency, we instead allocate multiple buffers per worker, enabling multiple packets to be in flight concurrently.

\noindent\textbf{Credit-based coordination.}
We implement a credit-based mechanism~\cite{kung1995creditfc,erpc19} to prevent buffer exhaustion at the receiver: the sender maintains a credit count, representing the available receive buffers at the remote node \numcircledmod{5}.
This count is decremented each time a packet is sent and replenished upon receiving acknowledgments (credit returns) from the remote node.
For example, if a remote worker has 128 receive buffers allocated, the initial credit for communication is set to 128.
Each packet transmission reduces this credit by one, and each acknowledgment restores it accordingly.
Since we have direct control over packet destinations via queue targeting (i.e., which specific thread each packet is delivered to), we can precisely track credit consumption and replenishment on a per-thread and node basis.
For instance, thread 0 on one node only talks to its buddy thread 0 on the other node.
Our PortPicker and the underlying awareness of the NIC architecture have enabled this per-thread communication pattern.
The allocation and initial assignment of receive buffers (credits) are determined during system startup.

\begin{figure}
    \centering
    \includegraphics[width=0.97\columnwidth]{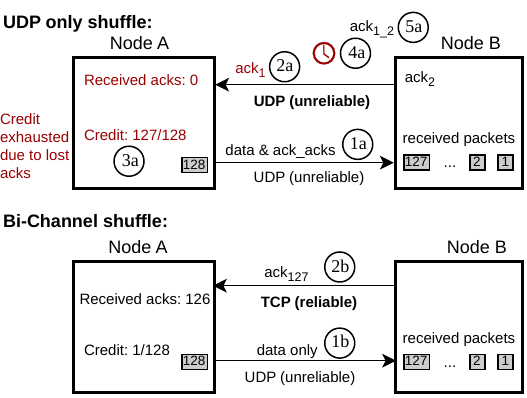}
    \caption{A pure UDP-based shuffle implementation (top) comes with a significant implementation complexity to ensure reliable data transmission (steps 1a-5a). The bi-channel approach (steps 1b-2b) combines the efficiency of UDP and the reliability of TCP (bottom).}
    \label{fig:bi_channel_shuffle}
  \end{figure}

\noindent\textbf{What can go wrong?}
Consider \Cref{fig:bi_channel_shuffle} (top), which illustrates nodes A and B communicating over UDP only.
Node A transmits data \numcircledmod{1a}, while node B returns a credit acknowledgment \numcircledmod{2a}.
If this acknowledgment is lost, the corresponding credit is not replenished.
At best, this merely reduces bandwidth utilization by reducing the number of packets in flight.
In the worst-case, repeated acknowledgment losses could lead to complete credit exhaustion \numcircledmod{3a}, causing node A to block indefinitely and halt progress.
One potential solution is for node A to send an additional acknowledgment, confirming receipt of node B's credit acknowledgment.
However, this "acknowledgment of acknowledgment" could also be lost, forcing node B to use a timeout \numcircledmod{4a} and periodically retransmit the original credit acknowledgment \numcircledmod{5a}.
Unfortunately, such a timeout-based retransmission can cause issues: suppose the original acknowledgment was successfully received, but only the acknowledgment-of-acknowledgment was lost.
In that case, retransmission might erroneously increment the credit twice, causing receiver buffers to underflow as credits exceed actual available buffers.
To address this duplication issue, we might introduce sequence numbers to identify and eliminate redundant retransmissions. 
However, another complexity emerges: acknowledgment packets themselves consume receive buffers, requiring additional buffer management.
Clearly, the complexity escalates, prompting an attentive reader to notice that we are effectively reimplementing fundamental TCP mechanisms, including flow control, buffer management, and sequence numbering.
This observation raises the question of whether such a reimplementation from scratch is necessary.

\noindent\textbf{Control path to the rescue.}
The solution provided by the bi-channel approach is straightforward: use the reliable TCP-based control channel to transmit credit acknowledgments.
Similarly, we use the TCP-based control channel also to re-transmit any outgoing packets that were not acknowledged by the receiver.
Because TCP inherently ensures reliability through established mechanisms such as flow control and retransmission, our implementation is significantly simplified.
In summary, the fast channel manages the high-throughput data path \numcircledmod{1b}, while the control channel, using TCP, reliably handles credit acknowledgments and packet retransmissions \numcircledmod{2b} (cf. \Cref{fig:bi_channel_shuffle} bottom).

\subsection{Evaluation}
In the following, we evaluate the performance of the shuffle operator using the bi-channel communication paradigm.
Our experiments show that the bi-channel approach offers a significant efficiency advantage over the traditional TCP stack.
Furthermore, we show that the bi-channel approach scales well with cluster size and achieves higher throughput than the kernel TCP stack for small table sizes.
  
\begin{figure}
      \centering
      \includegraphics[width=0.99\columnwidth]{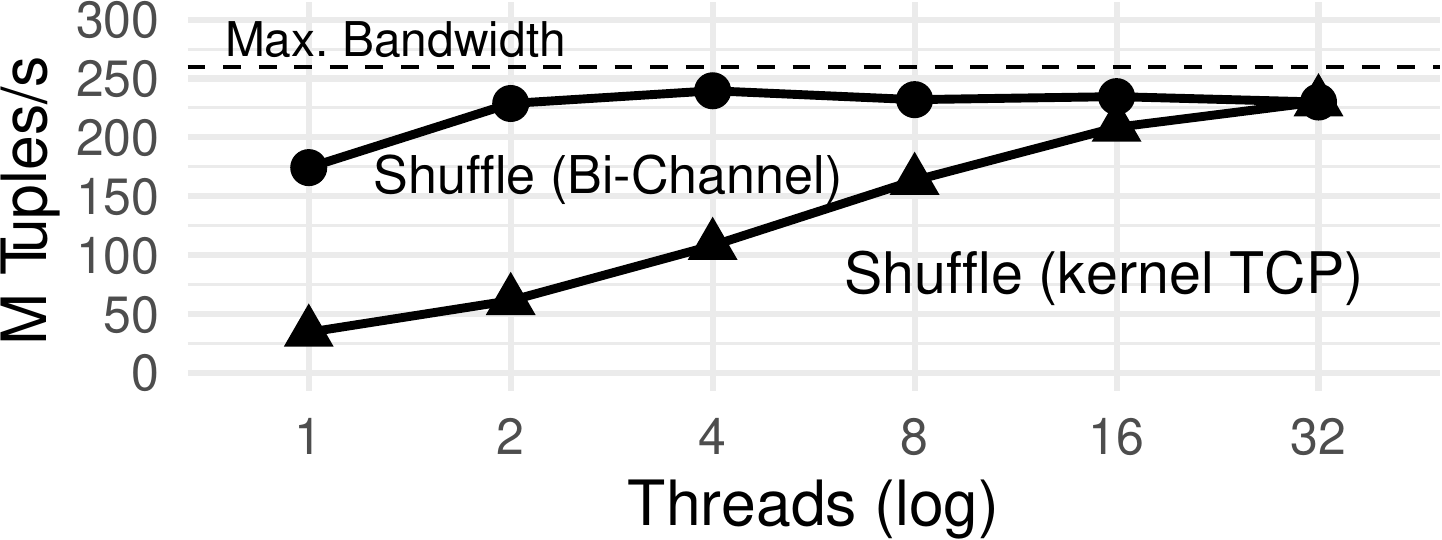}
      \caption{Shuffle operator benchmark results showing the impact of an increasing number of threads in the networking layer on tuple throughput. Threads are always fully-utilized. Each node shuffles 256 GiB.}
      \label{fig:shuffle_scalability}
  \end{figure}

\noindent\textbf{Experimental setup.}
The shuffle is performed by scanning data from an in-memory table in columnar format with 16 columns, each 8 bytes wide.
The data is partitioned by hashing one input column with a murmur hash function \cite{murmurhash}.
The result data is gathered in row-based format, simulating the gather phase of a hash join operation.

The kernel TCP implementation uses the io\_uring interface with submission queue poll mode.
For sending data, we use io\_uring's zero-copy send interface~\cite{iouringzc} to avoid main memory bandwidth becoming a bottleneck.
We use 256 TCP connections and a buffer size of 256 KiB, as this setup proved to be the most efficient.
On submission, we set the \textit{IOSQE\_ASYNC} flag to indicate that the operation should be executed by a kernel worker thread.
Our benchmark scales the kernel worker threads, while the networking layer uses a single thread for submitting tasks and processing completions.

We conduct the experiments on the c7gn.16xlarge instance type with 200 Gbps network bandwidth in AWS.
The instances are running Ubuntu 24.04 with Linux kernel version 6.8.

\noindent\textbf{Scalability.}
In the first benchmark, we evaluate the performance of the shuffle operator on a four-node cluster, shuffling 256 GiB of data.
Since the main memory size of the c7gn.16xlarge instance is limited to 128 GiB, we iterate multiple times over a table with 64 GiB of data.
The query processing layer uses 8 threads, producing enough remote tuples to saturate the network bandwidth.

\Cref{fig:shuffle_scalability} shows the effect of increasing the number of fully-utilized threads in the networking layer on the throughput of the shuffle operator.
The bi-channel approach achieves a tuple throughput of 175 M tuples per second with a single thread in the networking layer.
The kernel TCP stack, on the other hand, achieves less than 50 M tuples per second with a single networking-layer thread.

Scaling the bi-channel approach to four threads maximizes the throughput at 240 M tuples per second, reaching over 90\% of the theoretical maximum of 260 M tuples per second.
The kernel TCP stack, on the other hand, only reaches 230 M tuples per second even when scaling up to 32 kernel worker threads.
At this point, the TCP shuffle operator uses more than 50\% of all CPU resources exclusively for network communication.

\begin{figure}
    \centering
      \includegraphics[width=0.99\columnwidth]{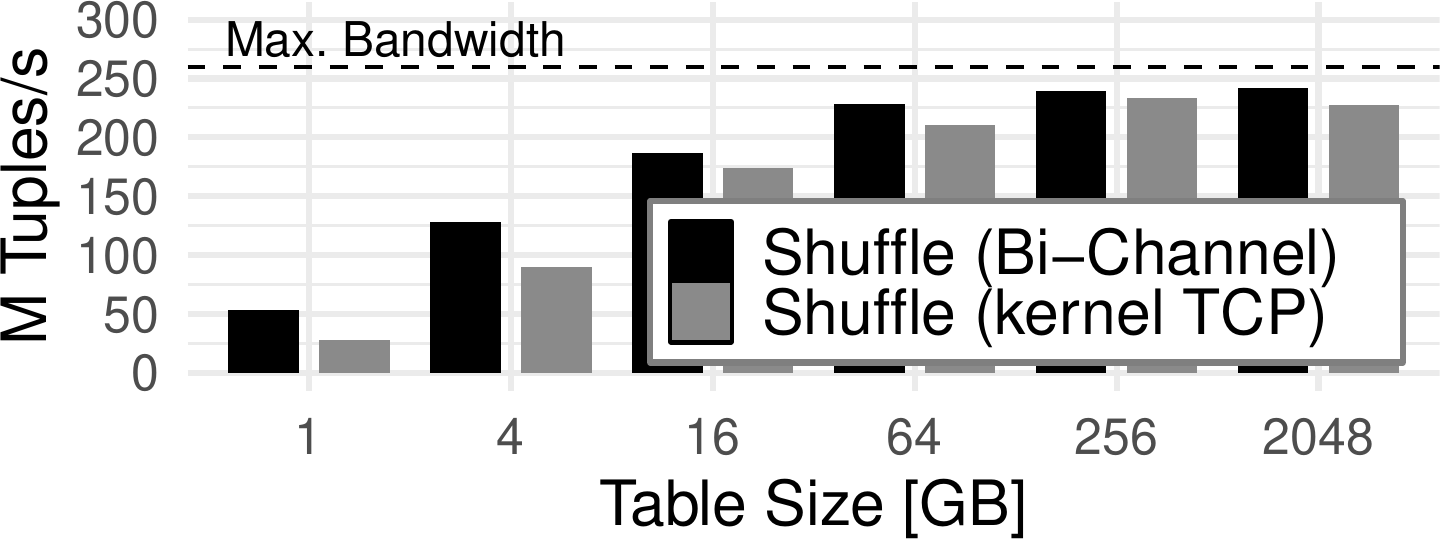}
      \caption{Shuffle operator benchmark results for increasing table size. Each node uses 4 threads (Bi-Channel) and 32 threads (kernel TCP).}
      \label{fig:shuffle_table_size}
\end{figure}

\noindent\textbf{Table size.}
In the second benchmark, we evaluate shuffle performance across a range of table sizes.
\Cref{fig:shuffle_table_size} compares the kernel-based shuffle implementation with the bi-channel variant for table sizes ranging from 1~GB to 2~TB.

For 1~GB tables, the bi-channel approach is twice as fast.
This result shows that the bi-channel paradigm also benefits less data-intensive operations such as distributed aggregations.
Whereas the kernel-based shuffle uses all available cores, the bi-channel implementation runs on only four.
This advantage stems from TCP overheads, including congestion control, which requires shuffling more data to reach peak throughput.
The performance gap begins to narrow at around 16~GB, where the larger data volume gives TCP enough time to ramp up and fully utilize the available bandwidth.
Although small tables may appear less relevant given modern data volumes, analyses of the Snowset~\cite{DBLP:conf/nsdi/VuppalapatiMATM20}, Redset~\cite{DBLP:journals/pvldb/RenenHPVDNLSKK24}, and trace-derived workload synthesis~\cite{DBLP:journals/corr/abs-2511-13059} show that nearly 90\% of tables are smaller than 1~GB, underscoring the importance of performance for small data sizes.

\section{Use Case 2: Low-Latency Transactional Systems}
\label{section:kvstore}
While the shuffle use case focused on bandwidth, this section demonstrates how the bi-channel paradigm benefits latency-critical applications.
We present a replicated key-value store that achieves very high message rates with predictable low latency.

\subsection{Anatomy}
\noindent\textbf{Replication use case.}
Modern distributed systems increasingly depend on replicated key-value stores for state management, session handling, and real-time feature storage.
Various replication approaches exist, including consensus algorithms such as Raft and Paxos.
In this use case, we focus on quorum-based replication, where a client sends key-value updates to a designated primary.
The primary replicates updates to multiple secondary nodes and waits for majority confirmation before responding to the client.

\noindent\textbf{Communication characteristics.}
In this use case, the main goal is to minimize replication latency.
While client communication generally remains beyond our direct control and commonly relies on TCP, internal replication presents an excellent opportunity for optimization.
Given that the primary node communicates with multiple secondary nodes, achieving high message rates with low latency is critical.
For instance, if clients generate one million requests per second and each request is replicated across five secondary nodes, the primary node must process a total of five million messages per second.
As a result, the effective incoming message rate increases proportionally to the replication factor.
Preserving the order of independent key updates is unnecessary; however, maintaining it for dependent updates is important.

\subsection{Bi-Channel Replicated KV-Store}
\label{subsec:bi-channel_replicated_kv-store}
We now demonstrate how the bi-channel paradigm can be applied to implement a replicated key-value store.
We highlight the flexibility of the bi-channel approach and emphasize that achieving optimal performance requires deliberate co-design of system requirements and the bi-channel implementation.

\begin{figure}
  \centering
  \includegraphics[width=0.97\columnwidth]{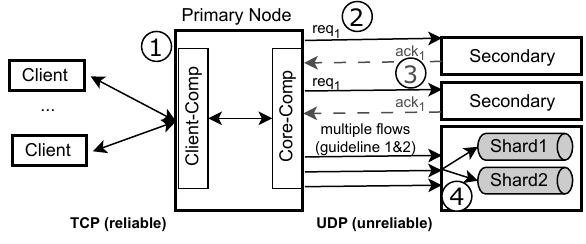}
  \caption{The bi-channel paradigm applied for KV-Store replication: Our approach combines a reliable TCP channel for client-facing communication with a low-latency UDP channel for data replication within a cluster.}
  \label{fig:kv-store-bi-channel}
\end{figure}

\noindent\textbf{System overview.}
\Cref{fig:kv-store-bi-channel} provides a high-level overview of the system architecture.
The primary node consists of a client-facing and a core component that processes key-value updates and manages replication.
Incoming requests \numcircledmod{1} are first handled by the client-facing component and then forwarded to the core component.
The core component replicates these requests to secondary nodes \numcircledmod{2}, which apply the updates to their local state and acknowledge receipt back to the primary \numcircledmod{3}.
Once the primary receives acknowledgments from a majority of secondaries, it applies the update to its local hash table and responds to the client's request.

\noindent\textbf{Thread scalability.}
As illustrated in the figure, we adopt a sharded architecture \numcircledmod{4}, commonly used in key-value stores.
In this design, dedicated worker threads handle client-facing requests and forward them to the appropriate shard-specific core workers via internal queues.
This separation ensures that each core worker can process updates independently, minimizing contention.
We carefully follow the guidelines to avoid the flow bandwidth limit and to choose appropriate ports to utilize the NIC, as discussed in \Cref{sec:cloud_nic}.
As in the shuffle use case, we leverage our knowledge of the specific worker handling each replication request.
This approach allows us to route replication requests directly to the corresponding shard on secondary nodes, avoiding contention points, e.g., mutexes.

\noindent\textbf{Receive Buffer Management.}
As in the shuffle use case, we employ a credit-based mechanism on the slow path to manage receive buffers.
However, unlike shuffle, we decouple replication acknowledgments from credit-based flow control messages.
Credits are sent asynchronously in batches over the control path, and we register enough receive buffers to sustain high throughput.
Replication acknowledgments, by contrast, are handled directly on the hot path, as they are latency-critical: the primary must wait for a quorum of these acknowledgments before confirming completion to the client.

\begin{figure}
      \centering
      \includegraphics[width=0.99\columnwidth]{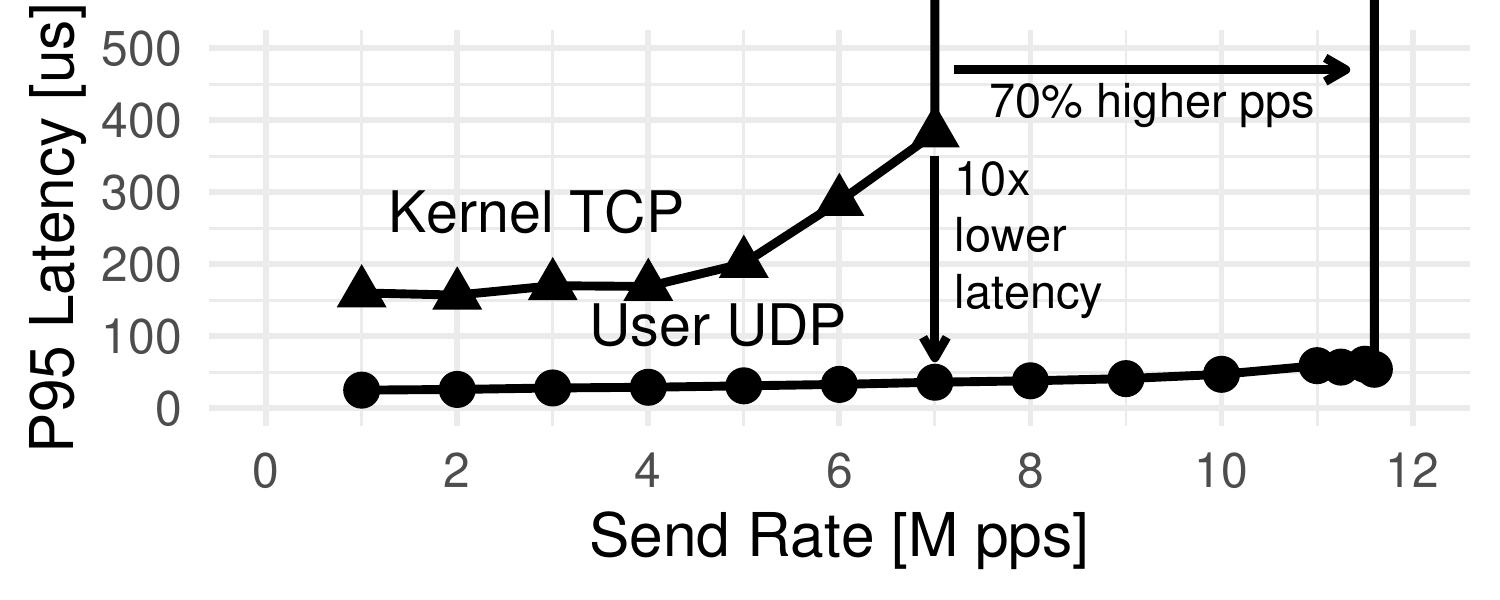}
      \caption{Impact of an increasing send rate on 95th percentile reply latency between two c7gn.16xlarge instances for Kernel TCP and User UDP with 64-byte messages.}
      \label{fig:kv-microbenchmarks}
\end{figure}

\noindent\textbf{Low-Latency quorum acknowledgments on the data path.}
To support this design, we rely on user-space UDP to send acknowledgments, enabling us to meet the strict latency requirements of the hot path.
\Cref{fig:kv-microbenchmarks} shows a microbenchmark comparing kernel-based TCP with user-space UDP, highlighting the difference in 95th percentile latency at a fixed message rate.
We use small 64-byte packets to minimize the latency overhead of transmission.
User-space UDP consistently achieves predictable latencies below 100 microseconds up to the NIC's bi-directional limit of 12 million messages per second.
In contrast, kernel TCP consistently exceeds 150 microseconds in latency and delivers 70\% fewer messages per second at a peak with 10 $\times$ higher latency.
These results demonstrate that relying on kernel TCP for acknowledgments (as in the shuffle) would severely degrade latency and throughput -- an unacceptable trade-off for a latency-critical system.
We implement a minimal, tailored subset of TCP functionality to address this, ensuring reliable delivery without unnecessary overhead.
For instance, congestion control is not required, as the primary node already regulates the load on secondaries.
We also simplify protocol semantics by leveraging domain-specific knowledge: version numbers used for recovery are reused as lightweight sequence numbers to ensure consistency across replicas.
Packet loss is rare due to adequate buffer sizing and is handled via simple timeouts and retransmissions.
Moreover, when data is replicated across four out of five nodes, retransmissions to the remaining node can be safely skipped thanks to redundancy.

\noindent\textbf{Client communication on the control path.}
However, there are other opportunities for utilizing the slow path.
Client communication differs fundamentally from the internal replication communication patterns.
In a static environment, such as our primary and secondary node cluster, we can easily pre-allocate receiver buffers, precisely calculate credits, and maintain persistent connections.
In contrast, clients typically connect intermittently and unpredictably, making it difficult to anticipate the number and frequency of connections.
As a result, static rate-limiting approaches, which evenly distribute credits across cluster nodes, are inadequate for client communications.
Instead, the primary node must dynamically allocate credits to clients based on the number of clients and their individual sending rates.
This requirement closely resembles the flow control mechanisms built into the TCP protocol.
Handling an arbitrary number of clients transmitting at varying rates introduces complexity beyond managing communications within a static node cluster.
Therefore, for client interactions, we leverage the kernel-based TCP path to effectively address these dynamic communication challenges.

\subsection{Evaluation}
We next evaluate the performance of the KV-store using the bi-channel communication paradigm.
Our experiments show that the bi-channel network stack achieves higher packet rates at lower latencies and CPU usage than the kernel TCP stack.
The end-to-end benchmark shows that the bi-channel approach enables the KV-store to fully saturate the network interface.

\begin{figure}
    \centering
    \includegraphics[width=0.99\columnwidth]{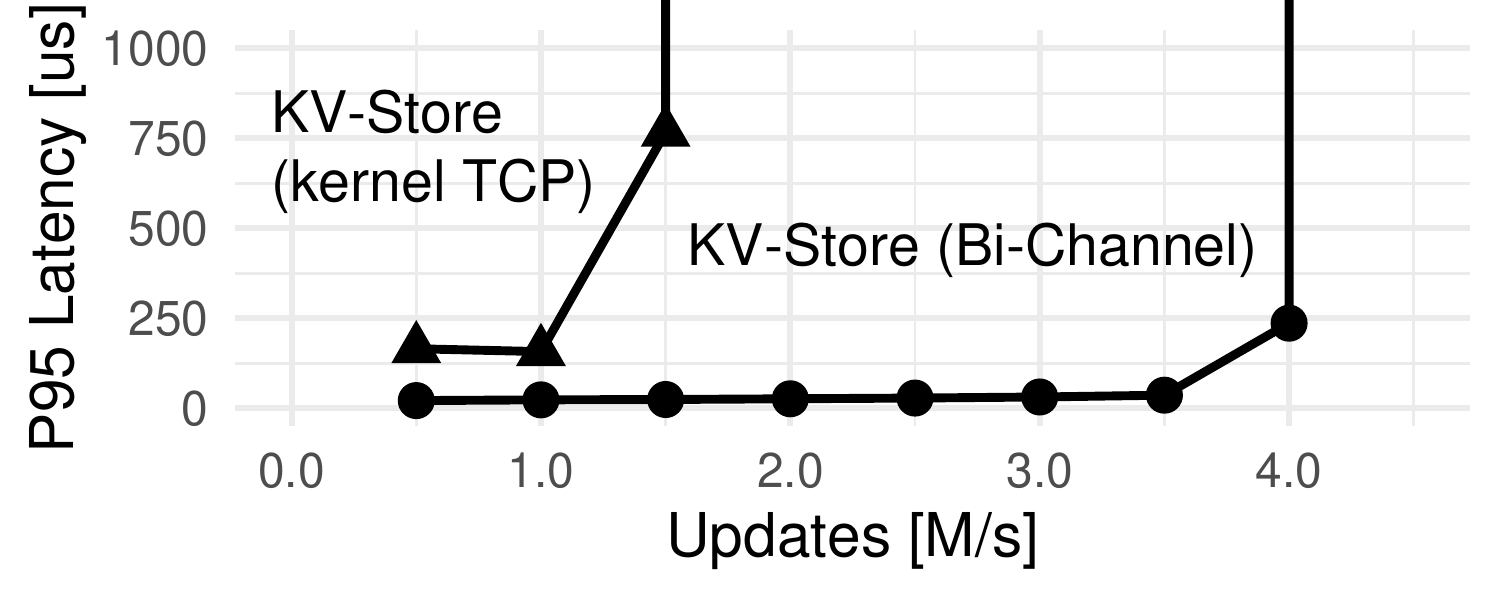}
    \caption{KV-Store replication benchmark with increasing update rate comparing the bi-channel approach to kernel TCP. The sender uses Poisson-distributed send timings.}
    \label{fig:kvstore-bi-channel-e2e_latency}
\end{figure}

\noindent\textbf{Methodology \& setup.}
We conduct the experiments in the same environment as in the previous use case. 
The kernel TCP implementation uses the io\_uring interface.
This time, compared to the shuffle benchmark, we avoid dispatching work to kernel worker threads by setting the defer task run option for improved latency.
We scale CPU resources by increasing the number of threads handling io\_uring submissions and completions.
Each thread uses a separate io\_uring instance.

We evaluate the performance of the KV-store replication process accelerated by the bi-channel paradigm.
We compare performance with a KV-store implementation that uses the kernel TCP stack instead of the bi-channel network stack.

\noindent\textbf{End-to-End latency measurement.}
The benchmark sends updates to the KV-store from a client at an increasing rate.
After receiving the update at the KV-store, the KV-store replicates its state and replies to the client once the update is durably stored.
We use small 64-byte updates to minimize the impact of transmission time on our latency measures.
We measure the end-to-end latency at the client.
This means that, for the KV-store with the bi-channel network stack, the total latency includes both the round-trip latency of TCP for client-server communication and the latency of the bi-channel network stack for replication.

\noindent\textbf{Results.}
\Cref{fig:kvstore-bi-channel-e2e_latency} shows the p95 latency of the two key-value store implementations under increasing update rates.
The bi-channel variant consistently achieves 100 microseconds lower p95 latency at low load than the kernel-based TCP version, and this gap widens as load increases.
The kernel TCP stack reaches a maximum throughput of 1.5 million updates per second.
In contrast, the bi-channel approach achieves up to 2.6 $\times$ higher throughput -- approximately 3.9 million updates per second -- while maintaining a p95 latency of around 250 microseconds.

At this peak rate, the primary node handles 24 million packets per second, saturating the NIC's packet-processing capacity.
We conclude that the bi-channel approach enables the key-value store to fully utilize modern network hardware, significantly outperforming the kernel TCP-based alternative.

\begin{figure}
  \centering
  \includegraphics[width=\columnwidth]{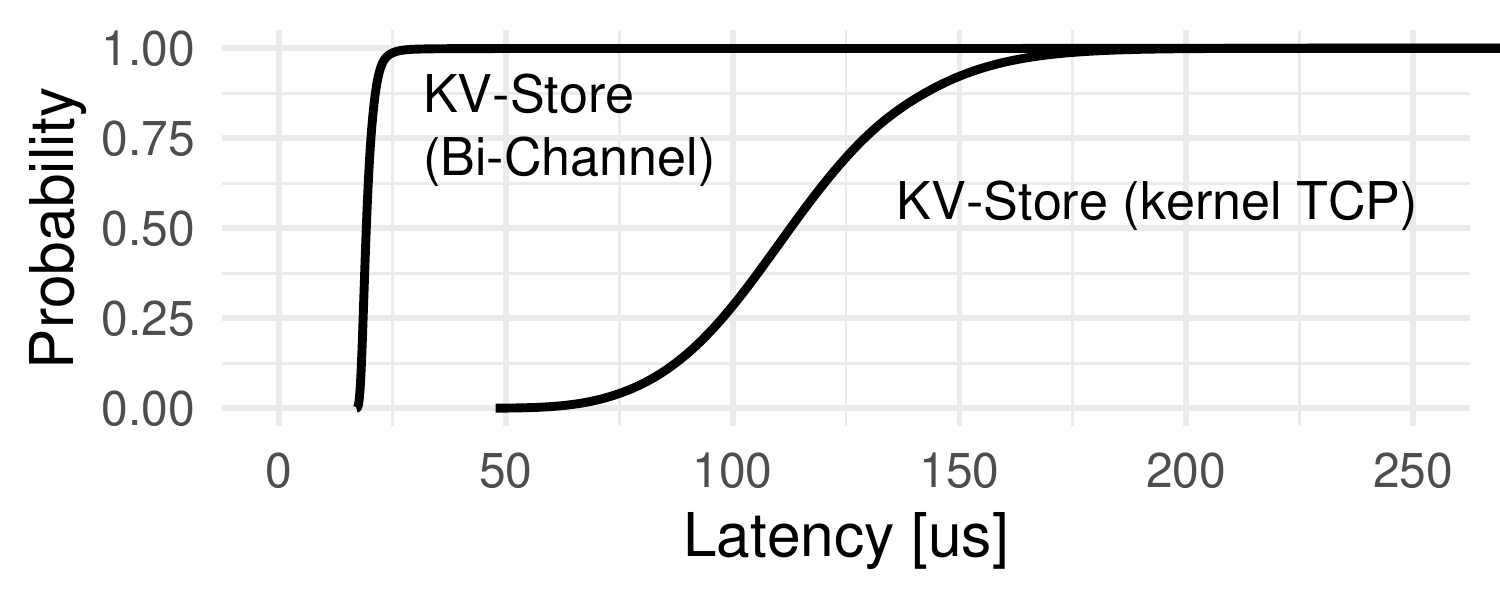}
  \caption{Cumulative distribution function for KV-Store replication with 1 M updates/s comparing the tail latencies of bi-channel and kernel TCP approach.}
  \label{fig:kvstore-latency-distribution}
\end{figure}

\noindent\textbf{Latency distribution.}
\Cref{fig:kvstore-latency-distribution} presents the cumulative distribution function (CDF) of latencies at a load of 1 million updates per second for both bi-channel and kernel TCP.
The y-axis represents cumulative probability, and the x-axis shows latency in microseconds.
We observe that bi-channel exhibits significantly more stable latency compared to kernel TCP.
For bi-channel, the difference between the minimum recorded latency and the 99th percentile latency is only 9 microseconds.
In contrast, for kernel TCP, this gap is approximately 140 microseconds; even the median latency deviates by around 80 microseconds from the minimum.
These results indicate that bi-channel offers substantially better predictability and stability, making it more suitable for scenarios requiring strict Service Level Agreements (SLAs).


\section{Related Work}\label{section:related_work}
\noindent\textbf{Hardware-supported kernel bypass.}
Several studies have explored RDMA~\cite{DBLP:journals/usenix-login/KaliaKA16, theendofslow} and EFA~\cite{DBLP:conf/damon/0001MLB22, DBLP:journals/micro/ShalevABS20}, networking technologies that bypass the kernel data path and expose low-latency communication to user space.
These solutions, however, require specialized hardware.
EFA, for example, is proprietary and currently available only on AWS, while RDMA is not widely available in public clouds.
In the context of database systems, extensive research has examined how to exploit such specialized networks.
Systems such as FaRM~\cite{FaRM1, FaRM2, FaRM3}, Herd~\cite{herd}, ScaleStore~\cite{DBLP:conf/sigmod/0001BL22}, and others~\cite{DBLP:conf/sigmod/LiDSN16, DBLP:conf/usenix/MitchellGL13, NAM, DBLP:journals/corr/abs-2207-03027, DBLP:conf/vldb/MakaitMR24} build on these approaches.
Other work has targeted specific operations, including distributed joins~\cite{rdmashuffle}, data-structure optimization~\cite{DBLP:conf/icdcs/XiaoWGL019, liu2025extendible, DBLP:conf/adms/WangYKSB20, wang2022sherman}, replication~\cite{10.1145/3587096, 286500}, and operator offloading~\cite{DBLP:conf/cidr/KorolijaKKTMA22}.

\noindent\textbf{User-space network stacks.}
Several systems move network processing into user space to reduce kernel overhead, including mTCP~\cite{DBLP:conf/nsdi/JeongWJJIHP14}, IX~\cite{DBLP:conf/osdi/BelayPKGKB14}, TAS~\cite{DBLP:conf/eurosys/KaufmannSPSKA19}, F-Stack~\cite{fstack}, LUNA~\cite{DBLP:conf/usenix/ZhuSXSFMCWWLYCL23}, and Seastar~\cite{seastar}.
These stacks are attractive when applications require TCP-compatible semantics, but building a general-purpose, production-ready stack remains challenging.
ScyllaDB~\cite{scylla} and Yellowbrick~\cite{cusack2024yellowbrick} are among the few production database systems that rely on a user-space networking stack, namely Seastar.
More recent DPDK-based systems, including the fast path used in our work, and cloud-oriented approaches such as Machnet~\cite{sanaee2025fastuserspacenetworkingrest}, show that user-space networking can substantially improve performance in cloud environments.

\noindent\textbf{Specialized communication abstractions.}
Specialized Remote Procedure Call (RPC) protocols have been developed for low-latency, small-message communication.
While some systems use RDMA~\cite{DBLP:conf/eurosys/SuZCGW17}, others rely on datagram-based designs~\cite{erpc19}.
Higher-level data-processing frameworks such as MPI~\cite{DBLP:journals/ijpp/LiuWP04} and DFI~\cite{DBLP:journals/sigmod/ThostrupSJZB22} abstract away transport details and provide structured communication models, such as bulk-synchronous parallelism.
These abstractions are effective when application logic matches their communication model.

\noindent\textbf{Transport offload and alternative transports.}
Recent work has accelerated TCP by offloading parts of the kernel network stack to FPGAs or SmartNICs~\cite{DBLP:conf/nsdi/MoonLJP20,DBLP:conf/nsdi/KimNGKYP23,DBLP:conf/nsdi/ShashidharaSKP22}.
A complementary line of work proposes alternative transport protocols for modern large-scale datacenter networks~\cite{DBLP:conf/conext/GaoNK0RS15, DBLP:conf/sigcomm/HandleyRAV0AW17, DBLP:conf/sigcomm/MontazeriLAO18, DBLP:journals/micro/ShalevABS20}.
Tux~\cite{DBLP:journals/pvldb/ZhouLYS25} recently explored a database-oriented kernel-bypass networking stack that combines a message-based transport protocol with DBMS-specific pushdown mechanisms.
These approaches are orthogonal to the bi-channel paradigm and complement it: when available in the target environment, they can serve as alternative implementations of the fast data channel instead of a DPDK/UDP-based fast path.

\noindent\textbf{Control/data-plane separation.}
The separation of control and data planes originates in networking research, where it laid the foundation for Software-Defined Networking (SDN)~\cite{DBLP:conf/sigcomm/CasadoFPLMS07}.
Beyond networking, this pattern has been widely adopted in distributed storage systems~\cite{DBLP:conf/sosp/GhemawatGL03,DBLP:conf/osdi/WeilBMLM06,DBLP:conf/mss/ShvachkoKRC10}.
Cloud database systems such as Snowflake~\cite{DBLP:conf/sigmod/DagevilleCZAABC16} apply a related principle by separating query planning and optimization from distributed query execution.
In contrast to this coarse-grained architectural separation, the bi-channel paradigm applies the idea at a finer granularity: it separates control and data paths within the networking layer of DBMS operators.
This systems-level interpretation is closer in spirit to kernel-bypass operating systems such as Arrakis~\cite{DBLP:conf/osdi/PeterLZPWKAR14} and Demikernel~\cite{DBLP:conf/sosp/ZhangRPONLMLSJP21}.


\section{Conclusion \& Future Work}
\label{section:conclusion}

We presented the \emph{bi-channel paradigm}, a new architectural principle for network-intensive database systems that separates communication into two coordinated planes: 
a high-performance, user-space data path and a reliable, kernel-based control path. 
This design enables systems to fully exploit modern NICs and cloud networking hardware without re-implementing TCP semantics in user space.

Our evaluation demonstrated that this paradigm combines performance and simplicity: 
a distributed shuffle operator saturates 200~Gbit/s using only three CPU cores, and a replicated key-value store sustains tens of millions of messages per second with predictable latency.
Beyond raw performance, the bi-channel approach yields a clearer division of concerns, simplifying the design of distributed operators.
We therefore view the bi-channel paradigm not merely as an optimization, but as a reusable systems pattern for database networking on emerging cloud hardware.

In this work, we chose user-space UDP for the fast path.
Future work will explore integrating the paradigm with kernel-native fast paths such as \texttt{AF\_XDP}, which can provide similar performance without external dependencies like DPDK.

\section{Artifacts}
\label{section:artifacts}

The artifacts for reproducing our experiments are available at:
https://github.com/GeorgKreuzmayr/bi-channel

The repository includes source code, experiment scripts, and reproduction instructions.

\balance
\begin{acks}
\noindent\includegraphics[width=2em]{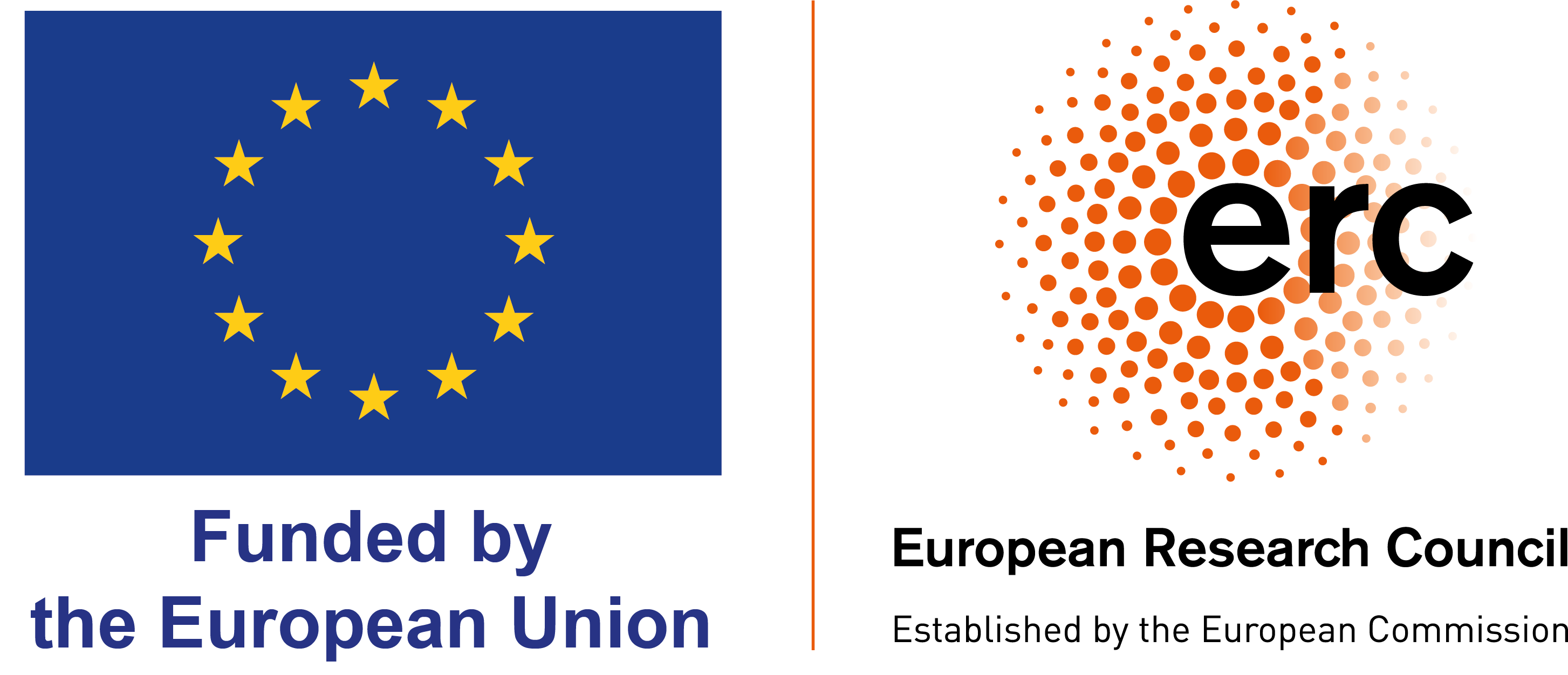} Funded/Co-funded by the European Union (ERC, CODAC, 101041375).
Views and opinions expressed are however those of the author(s) only and do not necessarily reflect those of the European Union or the European Research Council.
Neither the European Union nor the granting authority can be held responsible for them.
\end{acks}

\bibliographystyle{ACM-Reference-Format}
\bibliography{bib}


\end{document}